\pdfoutput=1

\documentclass[11pt,twoside,a4paper,cmspaper,final,collab]{cms-tdr}
\def\svnVersion{337691}\def\cmsCernNoTag{CERN-PH-EP/2013-037}\def\cmsCernDate{\today}\def\cmsMessage{Published in Physics Letters B as \href{http://dx.doi.org/10.1016/j.physletb.2016.03.060}{\doi{10.1016/j.physletb.2016.03.060.}}}

\begin{document}\cmsNoteHeader{TOP-12-033}

\hyphenation{had-ron-i-za-tion}
\hyphenation{cal-or-i-me-ter}
\hyphenation{de-vices}
\RCS$Revision: 326499 $
\RCS$HeadURL: svn+ssh://svn.cern.ch/reps/tdr2/papers/TOP-12-033/trunk/TOP-12-033.tex $
\RCS$Id: TOP-12-033.tex 326499 2016-02-18 14:37:05Z froscher $
\cmsNoteHeader{TOP-12-033}
\title{Inclusive and differential measurements of the \texorpdfstring{$\ttbar$}{t t-bar} charge asymmetry in \texorpdfstring{$\Pp\Pp$}{pp} collisions at \texorpdfstring{$\sqrt{s} = 8\TeV$}{sqrt(s) = 8 TeV}}

\date{\today}

\abstract{
  The \ttbar charge asymmetry is measured in proton-proton collisions at a centre-of-mass energy of $8\TeV$. The data, collected with the CMS experiment at the LHC, correspond to an integrated luminosity of 19.7\fbinv. Selected events contain an electron or a muon and four or more jets, where at least one jet is identified as originating from b-quark hadronization. The inclusive charge asymmetry is found to be $ 0.0010 \pm  0.0068\stat\pm 0.0037\syst$. In addition, differential charge asymmetries as a function of rapidity, transverse momentum, and invariant mass of the \ttbar system are studied. For the first time at the LHC, the measurements are also performed in a reduced fiducial phase space of top quark pair production, with an integrated result of $ -0.0035 \pm  0.0072\stat\pm 0.0031\syst$. All measurements are consistent within two standard deviations with zero asymmetry as well as with the predictions of the standard model.
  }

\hypersetup{%
pdfauthor={CMS Collaboration},%
pdftitle={Inclusive and differential measurements of the t t-bar charge asymmetry in pp collisions at sqrt(s) = 8 TeV},%
pdfsubject={CMS},%
pdfkeywords={CMS, physics, top quark, charge asymmetry}}

\maketitle

\section{Introduction}

The top quark offers an excellent opportunity to search for deviations from the standard model (SM), as its large mass makes it unique among all quarks. A possible hint for new physics in the top quark sector is the discrepancy between the measured \ttbar forward-backward asymmetry and the SM expectations, reported by the CDF~\cite{PhysRevD.83.112003, Aaltonen:2012it} and D0~\cite{PhysRevD.84.112005,Abazov:2014cca, PhysRevD.92.052007} collaborations at the Tevatron. Although this discrepancy has become smaller as the measurements and SM calculations~\cite{Czakon:2014xsa,Kidonakis:2015ona} have been refined, it has generated a number of theoretical explanations invoking contributions from physics beyond the SM (BSM).
These have in turn led to models based on axigluons or \zp bosons as mediators in the \ttbar production process.
An overview of the theoretical explanations can be found in Ref.~\cite{Aguilar-Saavedra:2014kpa} and references therein.

At hadron colliders top quark pairs are produced predominantly in the processes of gluon-gluon fusion and quark-antiquark annihilation.
At leading order (LO), the \ttbar production is symmetric with
respect to the exchange of the top quark and antiquark. At higher orders, QCD
radiative corrections to the $\cPq\cPaq \to \ttbar$ process induce an
asymmetry in the differential distributions of top quarks and antiquarks. The
interference between initial- and final-state radiation (ISR and FSR)
processes, as well as the interference between the Born and box diagrams,
generate a correlation between the direction of the top quark momentum and
that of the incoming quark~\cite{Kuhn:1998jr}. Similarly, the direction of the top antiquark momentum
is related to that of the incoming antiquark. These
processes induce a forward-backward asymmetry (\AFB) at the Tevatron $\Pp\Pap$
collider. The charge-symmetric pp collisions at the CERN LHC result in a
different effect. At the LHC, the larger average momentum fraction of the
valence quarks leads to an excess of top quarks produced in the forward and
backward directions, while the top antiquarks are produced more
centrally. This makes the difference in the absolute values of the rapidities\footnote{The rapidity is defined as $y = (1/2)\ln[(E+p_{z})/(E-p_{z})]$, where $E$ denotes the particle energy and $p_{z}$ its momentum component along the counterclockwise beam direction.}
of the top quark and antiquark, $\Delta \abs{y} = \abs{y_{\cPqt}} - \abs{y_{\cPaqt}}$, a suitable observable to measure the \ttbar charge asymmetry at the LHC experiments.
Using the sensitive variable, the charge asymmetry can be defined as
\begin{linenomath}
\begin{equation}
\label{eq:ac_def}
A_{\mathrm{C}} = \frac{N^{+} - N^{-}}{N^{+} + N^{-}},
\end{equation}
\end{linenomath}
where $N^{+}$ and $N^{-}$ represent the number of events with positive and negative values of $\Delta \abs{y}$, respectively.
Theoretical predictions for this observable are of order $1\%$ in the SM~\cite{Kuhn:2011ri,Bernreuther:2012sx}, but its sensitivity to new physics makes measurements of the effect interesting even when the precision is not high enough to establish the existence of the SM charge asymmetry.
Both the CMS and ATLAS collaborations have published results based on the data collected at a centre-of-mass energy $\sqrt{s}=7\TeV$, which are in agreement with the SM predictions~\cite{Chatrchyan:2012cxa,Aad:2013cea,Chatrchyan:2014yta,Aad:2015jfa}.

To shed light on the possible existence and the nature of new physics contributions, it is crucial to measure not only the inclusive asymmetry but also $A_{\mathrm{C}}$ as a function of variables magnifying the \ttbar charge asymmetry.
For this purpose Eq.~(\ref{eq:ac_def}) is modified to consider only events in a specific bin of the given variable.

In this letter, we present an inclusive measurement and three differential measurements of the \ttbar charge asymmetry. The three differential variables, which are each sensitive to a different contribution to the charge asymmetry, include the \ttbar system rapidity $\abs{y_{\ttbar}}$, its transverse momentum $\pt^{\ttbar}$, and its invariant mass $m_{\ttbar}$.
The measurements use the data collected with the CMS experiment in 2012 corresponding to an integrated luminosity of 19.7\fbinv at $\sqrt{s}=8\TeV$.

The variable $\abs{y_{\ttbar}}$ is sensitive to the ratio of the contributions from the $\cPq\cPaq$ and gg initial states to \ttbar production. The charge-symmetric gluon fusion process is dominant in the central region, while \ttbar production through $\cPq\cPaq$ annihilation mostly produces events with the \ttbar pair at larger rapidities, which implies an enhancement of the charge asymmetry with increasing $\abs{y_{\ttbar}}$~\cite{Kuhn:2011ri}.

The ratio of the positive and negative contributions to the overall asymmetry depends on $\pt^{\ttbar}$. In the SM the interference between the Born and the box diagrams leads to a positive contribution, while the interference between ISR and FSR results in a negative contribution. The presence of additional hard radiation implies, on average, a higher transverse momentum (\pt) of the \ttbar system. Consequently, in events with large values of $\pt^{\ttbar}$, the negative contribution from the ISR-FSR interference is enhanced~\cite{Kuhn:2011ri}.

The charge asymmetry is expected to depend on $m_{\ttbar}$ since the contribution of the $\cPq\cPaq$ initial state process is enhanced for larger values of this variable. It is also sensitive to BSM contributions; new heavy particles could be exchanged between initial quarks and antiquarks and contribute to the \ttbar production (see, \eg Ref.~\cite{Rodrigo:2010gm} and references therein). The amplitudes associated with these new contributions would interfere with those of the SM processes, and depending on the model they could lead to an increasing \ttbar charge asymmetry with increasing $m_{\ttbar}$.

Because only a part of the \ttbar phase space is experimentally accessible, measurements of the charge asymmetry that are to be compared to theoretical predictions necessarily include an extrapolation to a more well-defined phase space.
To this end a fiducial phase space is defined that emulates the restrictions of the measurable phase space while allowing for the calculation of theoretical predictions. This minimizes the need for extrapolation, which can be subject to unpredictable uncertainties if there are significant BSM contributions. An additional extrapolation to the full phase space of top quark pair production is provided as well, which allows for an easier comparison to the results of other measurements and theoretical calculations.

\section{CMS detector}

The central feature of the CMS apparatus is a superconducting solenoid of 6\unit{m} internal diameter, providing a magnetic field of 3.8\unit{T}.
Within the solenoid volume are a silicon pixel and strip tracker, a lead tungstate crystal electromagnetic calorimeter (ECAL), and a brass and scintillator hadron calorimeter, each composed of a barrel and two endcap sections. The inner tracker measures trajectories of charged particles within the pseudorapidity range $\abs{\eta} < 2.5$, while the calorimeters provide coverage up to $\abs{\eta} = 3.0$.
The pseudorapidity is defined as $\eta=-\ln (\tan\theta/2)$, with the polar angle $\theta$ being measured relative to the counterclockwise beam direction.
The ECAL has an energy resolution of 3\% or better for the range of electron energies relevant for this analysis. Extensive forward calorimetry complements the coverage provided by the barrel and endcap detectors. Muons are measured in the pseudorapidity range $\abs{\eta} < 2.4$ using gas-ionization detectors embedded in the steel flux-return yoke outside the solenoid. Matching muons to tracks measured in the silicon tracker results in a relative \pt resolution for muons with $20 <\pt < 100\GeV$ of 1.3--2.0\% in the barrel and better than 6\% in the endcaps. The \pt resolution in the barrel is better than 10\% for muons with \pt up to 1\TeV~\cite{Chatrchyan:2012xi}. A more detailed description of the CMS detector, together with a definition of the coordinate system used and the relevant kinematic variables, can be found in Ref.~\cite{JINST}.

\section{Simulated samples}
\label{sec:SimSamples}
For several steps of the measurement, samples of simulated events are used to model both the signal process and the background contributions arising from the production of single top quarks and vector bosons in association with jets (\PW+jets and \cPZ+jets).
An additional background contribution from QCD multijet events is modelled using a template derived from data; see Section~\ref{sec:bgest}.
Top quark pairs are produced with the next-to-leading order (NLO) generator \POWHEG, version 1.0~\cite{Alioli:2010xd,Nason:2004rx,Frixione:2007vw,Re:2010bp}, using the CT10~\cite{Lai:2010vv} parton distribution functions (PDF).
The electroweak production of single top quarks, in the $t$-channel and in association with a \PW~boson (\cPqt\PW-channel), is simulated using \POWHEG and the CTEQ6M PDF set~\cite{Pumplin:2002vw}.
The production of electroweak vector bosons in association with jets is simulated using \MADGRAPH, version 5.1.3.30~\cite{madgraph}, and the CTEQ6L1~\cite{Pumplin:2002vw} PDF set.

For the simulation of \ttbar and single top quark events the top quark mass is set to 172.5\GeV.
For all samples, \PYTHIA, version 6.426~\cite{Sjostrand:2006za}, is used for the description of the parton showering and hadronization.
The simulations include additional proton-proton interactions in the same bunch crossing (in-time pileup) and in earlier/later bunch crossings (out-of-time pileup) with the same frequency of occurrence as observed in the data.

Differential cross section measurements~\cite{Khachatryan:2015oqa}
have shown that the \pt spectrum of the top quarks in \ttbar events is significantly softer than the one generated by the used simulation programs.
To correct for this effect, the simulated \ttbar sample is reweighted according to scale factors derived from these measurements.

\section{Event selection}
\label{sec:EventSel}

The analysis uses \ttbar events in which one of the \PW~bosons from a top quark decay subsequently decays into an electron or muon and the corresponding neutrino, and the other \PW~boson decays into a pair of quarks.
We therefore select events containing one electron or muon and four or more jets, at least one of which is identified as originating from the hadronization of a bottom quark.
To be considered for the offline analysis, the events must pass a single-electron or a single-muon trigger with \pt thresholds of $27$ and $24\GeV$ for the electron and muon, respectively.

The particle-flow (PF) algorithm~\cite{PFT-09-001,CMS:2010byl} is used to reconstruct electrons, muons, and jets in the event.
The algorithm reconstructs and identifies each individual particle with an optimized combination of information from the various elements of the CMS detector. The reconstructed PF candidates are divided into five classes: electrons, muons, photons, charged hadrons, and neutral hadrons.

The primary vertex of the event~\cite{TRK-11-001} is identified as the reconstructed vertex with the highest sum of squared transverse momenta of the associated charged particles. For an event to be accepted, the primary vertex must satisfy criteria on its location within the detector volume, as well as on the quality of its reconstruction.

Electron candidates are required to have a transverse momentum larger than $30\GeV$ and be within $\abs{\eta} < 2.5$, excluding the transition region between the ECAL barrel and endcaps of $1.44 < \abs{\eta_{\mathrm{sc}}} < 1.57$ since the reconstruction of an electron object in this region is not optimal, where $\eta_{\mathrm{sc}}$ is the pseudorapidity of the electron candidate supercluster~\cite{Khachatryan:2015hwa}.
Furthermore, electron candidates are selected based on the value of a multivariate discriminant, which combines different variables related to calorimetry and tracking parameters, but also \pt and $\eta$ of the electron candidate. The electron definition also encompasses a conversion rejection method aimed at identifying electrons from photon conversions. Detailed information on the electron reconstruction can be found in Ref.~\cite{Khachatryan:2015hwa}.

Muons are required to have $\abs{\eta} < 2.1$ and $\pt > 26\GeV$, with further requirements on the quality of the muon reconstruction and the compatibility with the primary vertex of the event. The $\eta$ requirement reflects the coverage of the single-muon trigger. Details on the muon reconstruction can be found in Ref.~\cite{Chatrchyan:2012xi}.

Additionally, electron and muon candidates must be isolated.
The isolation is quantified by the variable $I_{\text{rel}}^{\ell}$, defined as the sum of reconstructed transverse momenta of nearby PF objects divided by the lepton transverse momentum ($\pt^{\ell}$), corrected for pileup effects~\cite{Khachatryan:2015hwa} using the effective area (in $\eta$-$\phi$ space) of the lepton and the energy density in the event.
Electrons and muons are required to have  $I_{\text{rel}}^{\ell} < 0.1$ and $< 0.12$, respectively, using isolation cones with radii of $0.3$ and $0.4$ in $\eta$-$\phi$ space.

Events with additional electrons and muons are vetoed. The lepton veto is based on a looser definition of the lepton identification.
In this definition, electrons must have $\pt > 20\GeV$, $\abs{\eta} < 2.5$ and $I_{\text{rel}}^{\ell} < 0.15$, while passing a loose criterion on the value of the multivariate discriminant.
Muons are required to have $\pt > 10\GeV$, $ \abs{\eta} < 2.5$, and $I_{\text{rel}}^{\ell} < 0.2$.

Jets are clustered from PF particles with the anti-\kt~\cite{antikt} algorithm with a distance parameter of 0.5. Charged hadrons identified as originating from pileup vertices are removed before clustering into jets.
Further corrections~\cite{Chatrchyan:2011ds} to the jet energy are applied, depending on jet $\eta$ and \pt, the jet area in $\eta$-$\phi$ space, and the median \pt density of the event.
The selected jets must lie within $\abs{\eta} < 2.5$ and are required to have $\pt > 30\GeV$.
The jet \pt resolution in data is approximately 10\% worse compared to simulations.
To account for this, the transverse momenta of jets in the simulated samples are smeared accordingly.
Finally, jets from the hadronization of \cPqb~quarks are identified using the medium working point of the combined secondary vertex algorithm~\cite{Chatrchyan:2012jua}. The b tag identification efficiency of this algorithm for b jets with $\pt > 30\GeV$ and $\abs{\eta} < 2.4$ varies between 60 and 70\%, while the misidentification rate for jets arising from light quarks or gluons is about 1\%~\cite{CMS:2013vea}.

With the applied event selection we find a total of 171\,121 events with an electron in the final state, hereafter referred to
as the electron+jets channel, and 192\,123 events in the muon+jets channel.

\section{Definition of a fiducial phase space}
\label{sec:fidSpaceDef}
Because of the offline event selection, only a subset of the events collected by the CMS detector is used in the analysis. To allow for a comparison of the measurements with the theoretical calculations, an extrapolation to a well-defined phase space needs to be performed.
The extrapolation relies on a correct modeling of the ratio of the number of events in the measured phase space to that in the extrapolated one; such a ratio, however, may be affected by new physics.
The simple approach, which is extrapolation to the \emph{full phase space} of \ttbar production, entails a large dependence on the model assumptions.

As an alternative, a \emph{fiducial phase space} is defined using generator-level selection criteria that mimic the reconstruction-level criteria applied during the nominal selection.
The ratio of the number of fiducial events to the number of reconstruction-level selected events, determined from simulation, is then applied to the data to estimate the distribution of an observable in the fiducial region.

Because of the physical and topological similarity of events in the selected and fiducial phase spaces, new physics contributions are expected to affect both in approximately the same way, leaving the ratio unchanged.
Thus this extrapolation to the fiducial phase space is nearly model-independent.
It should be noted that the desired model-independence is achieved by using a purely multiplicative correction; a prior subtraction of non-fiducial \ttbar events in the selected phase space would require a larger reliance on the model assumptions, as there would be no cancellation of uncertainties.

Jets of generated particles in simulated events are used to emulate the selection steps acting on reconstructed jets.
Hadron-level particles are clustered into jets using the anti-\kt algorithm with a distance parameter of 0.5, as used for the reconstructed jets.
The clustering includes charged leptons and neutrinos, except those created in the leptonic decay of \PW~bosons originating from top quarks. It should be noted that the selection criteria for charged leptons are applied only to leptons originating from top quark decays.

Using these objects the following selection requirements are applied.
The event needs to contain exactly one electron (or muon) with $\pt > 30 \,(26)\GeV$ and $ \abs{\eta} < 2.5 \,(2.1)$.
Any event that contains an additional electron (or muon) with $\pt > 20 \,(10)\GeV$ and $ \abs{\eta} < 2.5$ is rejected.
At least four generator-level jets with $\pt > 30\GeV$, $ \abs{\eta} < 2.5$ are required. The event is rejected if the axes of any such jets have an angular separation of $\Delta R < 0.4$ to the lepton, where $\Delta R = \sqrt{(\Delta \eta)^2 + (\Delta \phi)^2}$ is calculated using the differences in the azimuthal angles $\phi$ and pseudorapidities $\eta$. This criterion serves as an emulation of the lepton isolation criteria, which use similar radii and are hard to implement for theoretical calculations.

The fiducial region contains about 10\% of the events of the full phase space.
Roughly 50\% of the events in the fiducial region pass the selection outlined in Section~\ref{sec:EventSel}, with the largest inefficiencies caused by the lepton selection and trigger requirements.
In comparison, only 1.5\% of the events outside the fiducial region fulfil the event selection criteria, making up about 20\% of the selected events.

\section{Estimation of background contributions}
\label{sec:bgest}
For the estimation of the background contributions we make use of the discriminating power of the transverse mass of the \PW~boson, $m_\mathrm{T}^\PW$, and of $M_3$, the invariant mass of the combination of three jets that corresponds to the largest vectorially summed \pt.
This estimation is necessary for the subtraction of the background contributions of the measured data, as described in Section~\ref{sec:Measurement}.
The $m_\mathrm{T}^\PW$ variable is calculated from the transverse momentum of the charged lepton $\ptvec^\ell$ and the missing transverse momentum vector \ptvecmiss.
The latter is defined as the \pt imbalance of the reconstructed PF objects, taking into account the propagation of jet energy corrections to this observable.
Its magnitude is referred to as \ETmiss.
Neglecting the lepton masses, $m_\mathrm{T}^\PW$ is defined as
\begin{linenomath}
\begin{equation}
m_\mathrm{T}^\PW = \sqrt{2(\ETmiss \pt^{\ell} - \ptvecmiss \cdot {\vec p}_{\mathrm{T}}^{\ell})} .
\end{equation}
\end{linenomath}
The background estimation is made with a binned maximum likelihood fit for each lepton channel.
A simultaneous fit in $m_\mathrm{T}^\PW$ and $M_3$ is performed in two disjoint data samples, corresponding to $m_\mathrm{T}^\PW < 50\GeV$ and $>$50\GeV. The $m_\mathrm{T}^\PW$ distribution is fitted in the low-$m_\mathrm{T}^\PW$ sample, which is rich in QCD multijet events and yields a good discrimination between the QCD multijet process and processes containing a genuine \PW~boson. The distribution of $M_3$ is not as dependent on the choice of event sample; it is fitted in the complementary high-$m_\mathrm{T}^\PW$ sample to avoid using the same events for both fits.

For the \ttbar, \PW+jets, \cPZ+jets, and single top quark processes, simulated samples are used to model the shapes of the $m_\mathrm{T}^\PW$ and $M_3$ distributions.
The contribution from multijet background events is estimated from data control samples containing nonisolated or poorly identified leptons.
Rate constraints corresponding to Gaussian uncertainties of 20\% are introduced into the likelihood function for the \cPZ+jets and single top quark processes according to the respective NLO cross sections, while the rates of the other processes are free parameters of the fit. The width of the constraints is motivated by the uncertainties of measurements and theoretical calculations of the corresponding cross sections~\cite{Khachatryan:2014iya,Chatrchyan:2014tua,Khachatryan:2014zya,Chatrchyan:2014dha,Kidonakis:2012db}. A detailed description of the fitting procedure can be found in Ref.~\cite{Chatrchyan:2011hk}.

\begin{figure*}[tp!]
 \centering
   \includegraphics[width=0.4\textwidth]{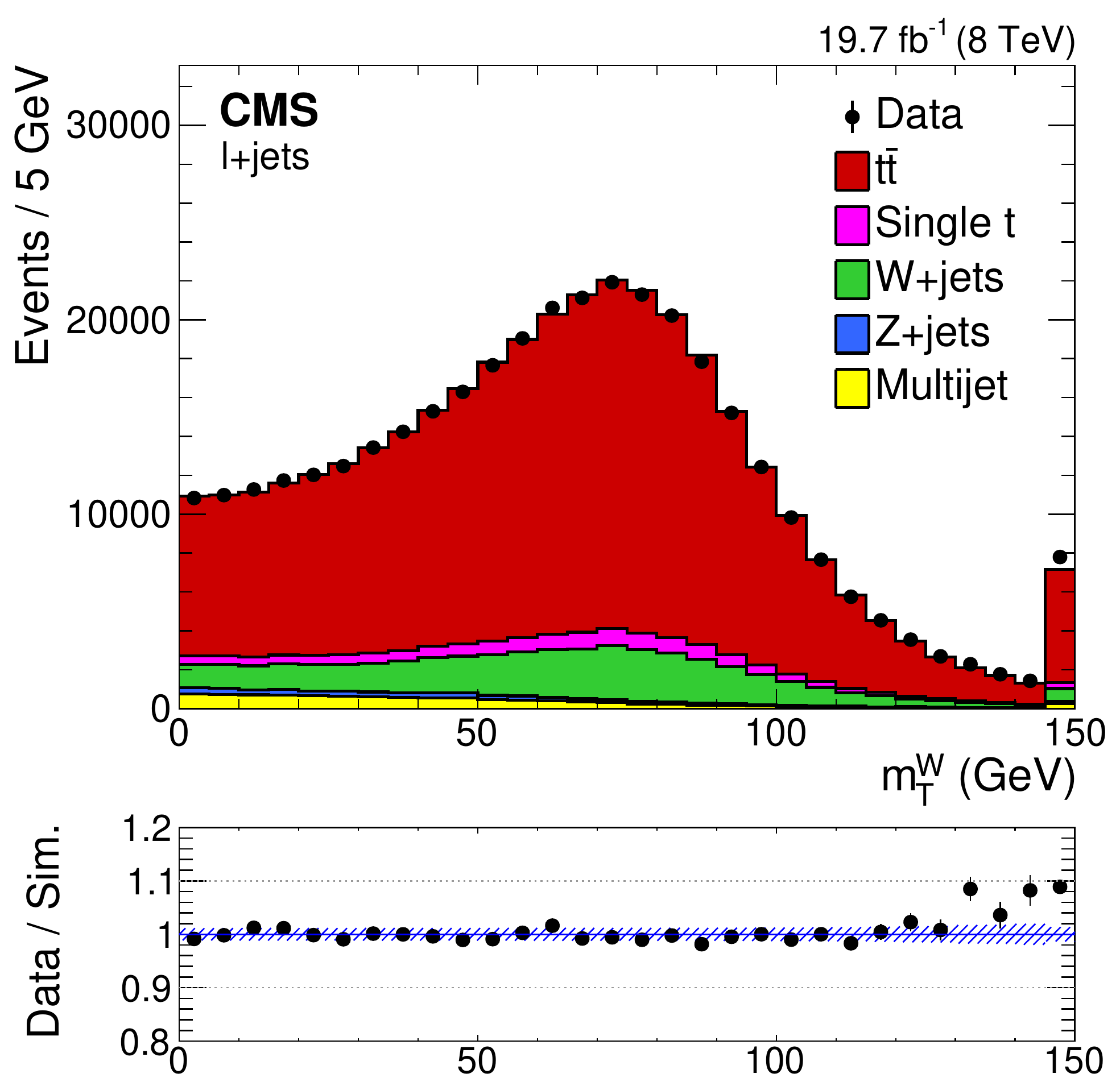}
   \includegraphics[width=0.4\textwidth]{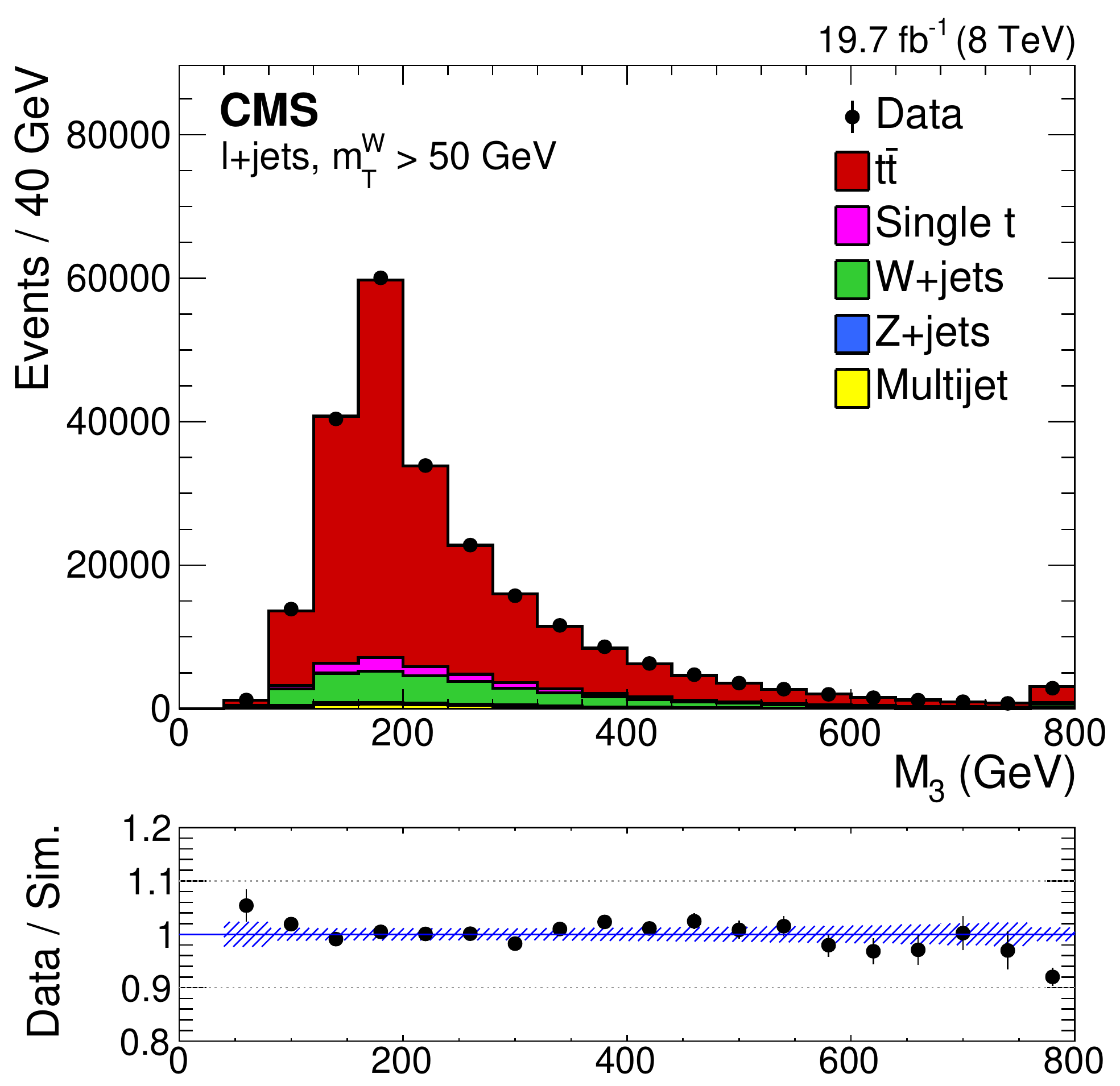}
     \caption{Comparison of the combined lepton+jets data with simulated contributions for the distributions in $m_\mathrm{T}^\PW$ and $M_3$. The simulated signal and background contributions are normalized to the results of the fits in Table~\ref{tab:FitResults}. The last bin in each histogram includes the overflow values. Additionally, the ratio of the data to the sum of the simulated contributions is shown, with the statistical uncertainties of the simulated contributions (including the uncertainties in the fit) indicated by the blue hatched region.}
 \label{fig:DataMCfitresult}
\end{figure*}

Table~\ref{tab:FitResults} summarizes the results of the fits. Figure~\ref{fig:DataMCfitresult} shows the two variables used for the estimation of the background, with the individual simulated contributions normalized to the results from the fit.

\begin{table*}[t]
 \topcaption{\label{tab:FitResults}Number of events for background and $\ttbar$ contributions from fits to data, along with their statistical uncertainties. The correlations between the individual values have been taken into account for the determination of the uncertainty on the total background yield. The uncertainties quoted for the single top quark and \cPZ+jets backgrounds are driven by the constraints used as inputs for the likelihood fit.}
\centering
    \begin{tabular}{lcc} \hline
    Process                          & Electron+jets                          &       Muon+jets            \\ \hline
    Single top quark ($t$ + \cPqt\PW)      & $ \hphantom{00\,}7016 \pm  1328 $          &    $ \hphantom{\,00}7302 \pm 1663 $   \\
    \PW+jets                         & $ \hphantom{0}22\,508 \pm 1460$            &    $ \hphantom{0}20\,522 \pm 1606 $           \\
    \cPZ+jets                        & $ \hphantom{00\,}2345 \pm 510\hphantom{0}$ &    $ \hphantom{\,00}2046 \pm 415\hphantom{0}$ \\
    QCD multijet                       & $ \hphantom{00\,}6136 \pm 1201$            &    $ \hphantom{\,00}4199 \pm 588\hphantom{0}$ \\
    \hline
    Total background                  &$\hphantom{0}38\,005 \pm 1491 $ &$\hphantom{0}34\,096 \pm 1495 $ \\
    $\ttbar$                          &$133\,130 \pm 1521$ & $158\,058 \pm 1538 $ \\
    \hline

    Observed data                     & 171\,121 & 192\,123 \\
    \hline
 \end{tabular}

\end{table*}

\section{Measurement of the \texorpdfstring{\ttbar}{t-tbar} charge asymmetry}
\label{sec:Measurement}

The measurement of the \ttbar charge asymmetry is based on the reconstructed four-momenta of the t and $\mathrm{\bar{t}}$ quarks in each event. We reconstruct the leptonically decaying \PW~boson from $\ptvec^{\ell}$ and \ptvecmiss and associate the measured jets in the event with quarks in the \ttbar decay chain.
The association is done using a likelihood criterion based on the \cPqb~tagging discriminator values of the jets and the corresponding reconstructed masses of the top quarks and \PW~bosons.
The reconstruction procedure is described in detail in Ref.~\cite{Chatrchyan:2011hk}.

The reconstructed top quark and antiquark four-momenta are used to obtain the inclusive and differential distributions of $\Delta\abs{y}$, and the charge asymmetry is calculated from the number of entries with $\Delta\abs{y} > 0$ and $< 0$.
In case of the differential measurements, the asymmetries are calculated separately for the different bins in the kinematic variable $V_{i}$, where $V_{i}$ is either $\abs{y_{\ttbar}}$, $\pt^{\ttbar}$, or $m_{\ttbar}$.

To allow for a comparison of the resulting asymmetry and the predictions from theory, the reconstructed distributions of $\Delta\abs{y}$ and the three kinematic variables are corrected for background contributions, resolution, and selection efficiency.

In the first correction step, the distributions of background processes, as used in Section~\ref{sec:bgest}, are normalized to the estimated rates (see Table~\ref{tab:FitResults}) and subtracted from the data, assuming Gaussian uncertainties in the background rates as well as in statistical fluctuations in the background templates. The correlations among the individual background rates are taken into account.

\begin{figure*}[thb]
 \centering
   \includegraphics[width=0.49\textwidth]{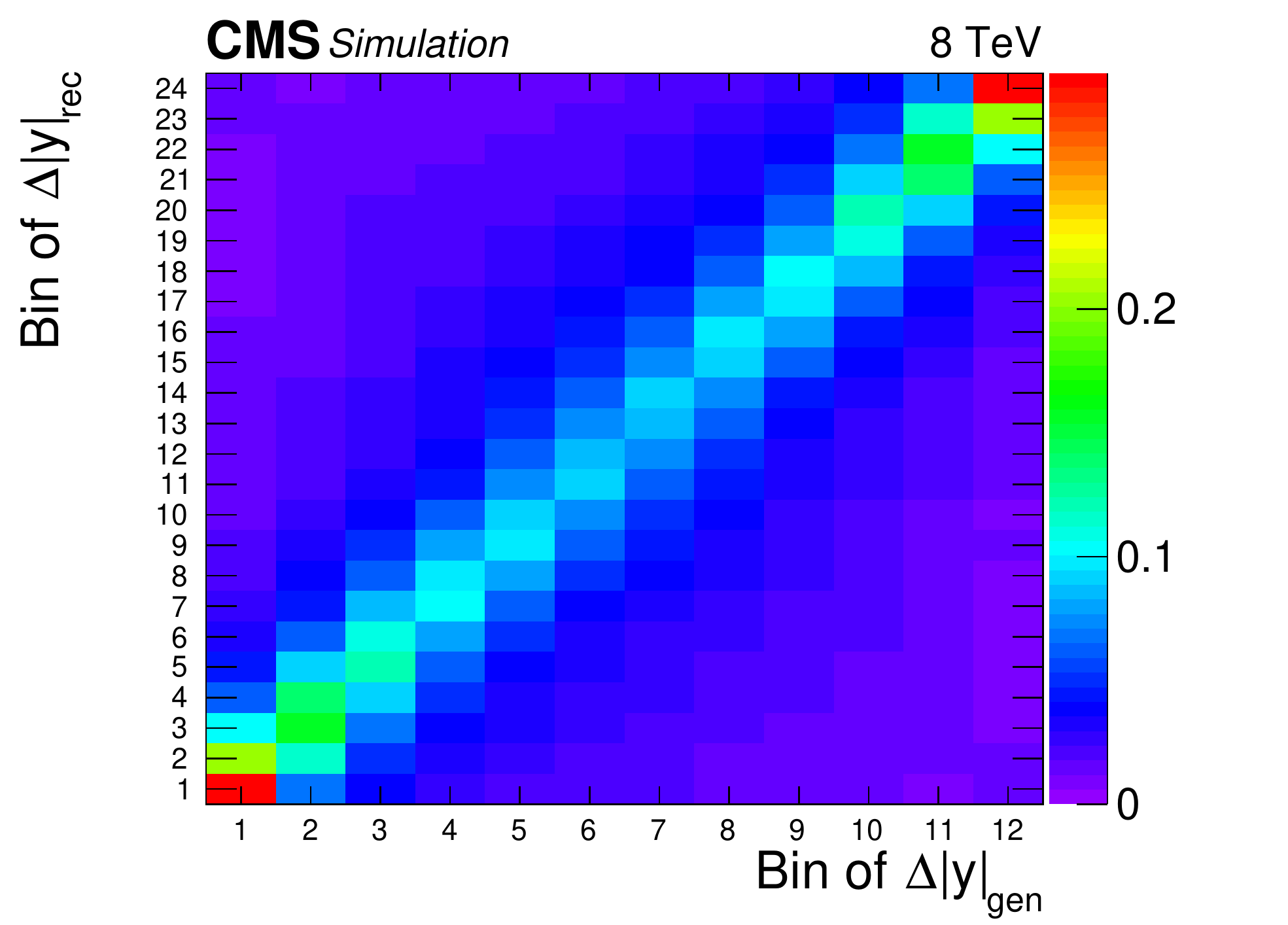}
   \includegraphics[width=0.49\textwidth]{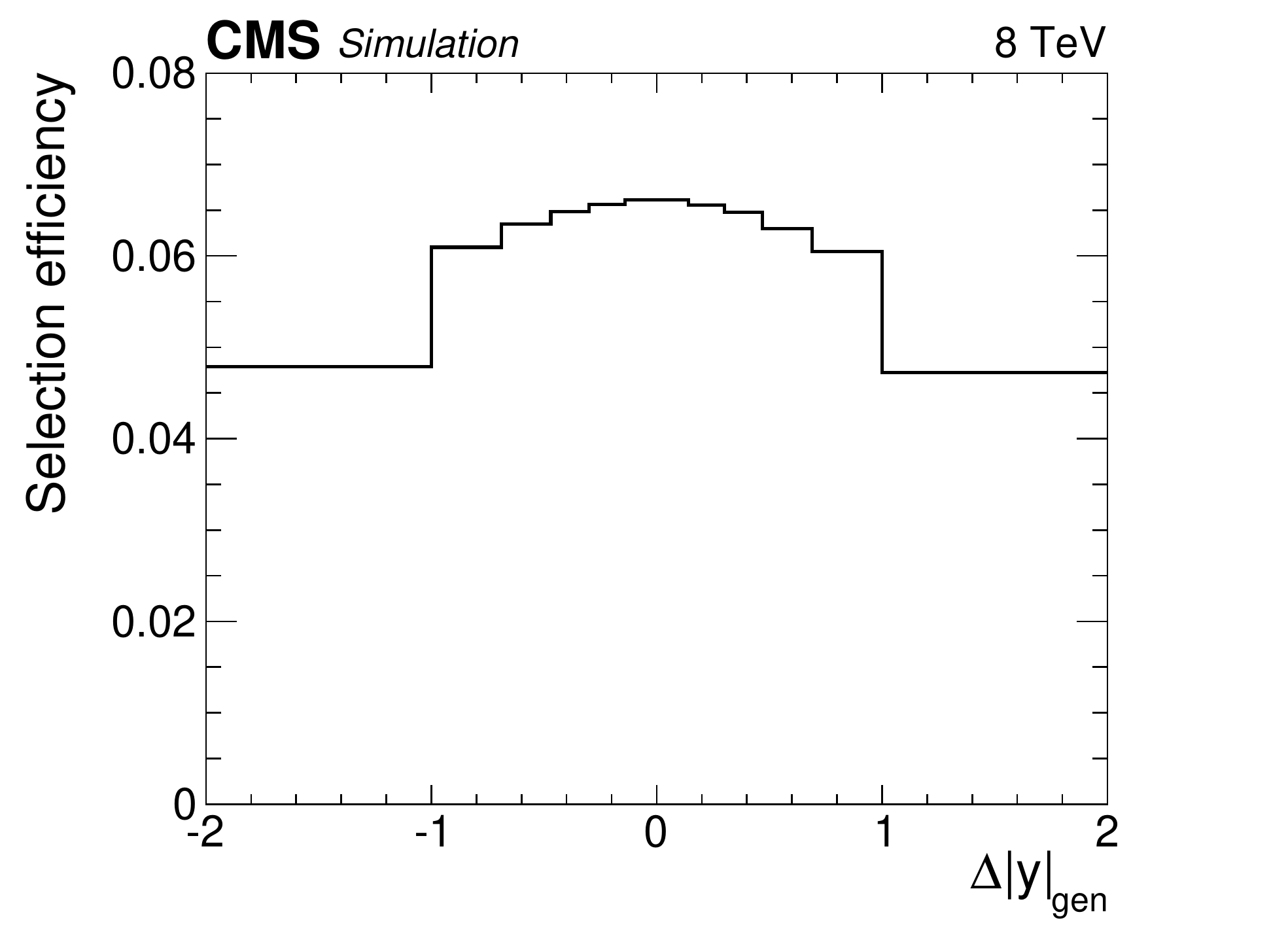}
   \includegraphics[width=0.49\textwidth]{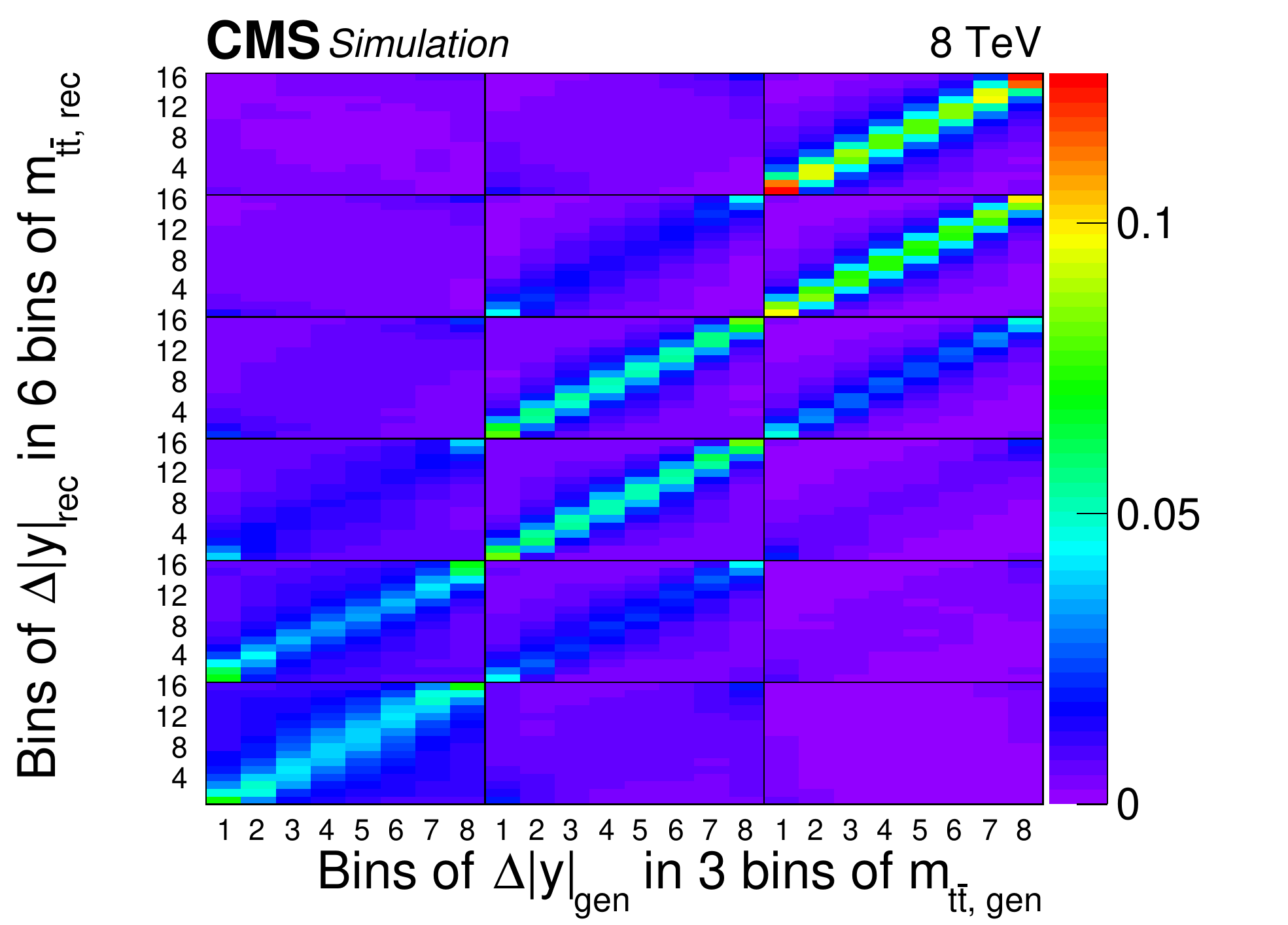}
   \includegraphics[width=0.49\textwidth]{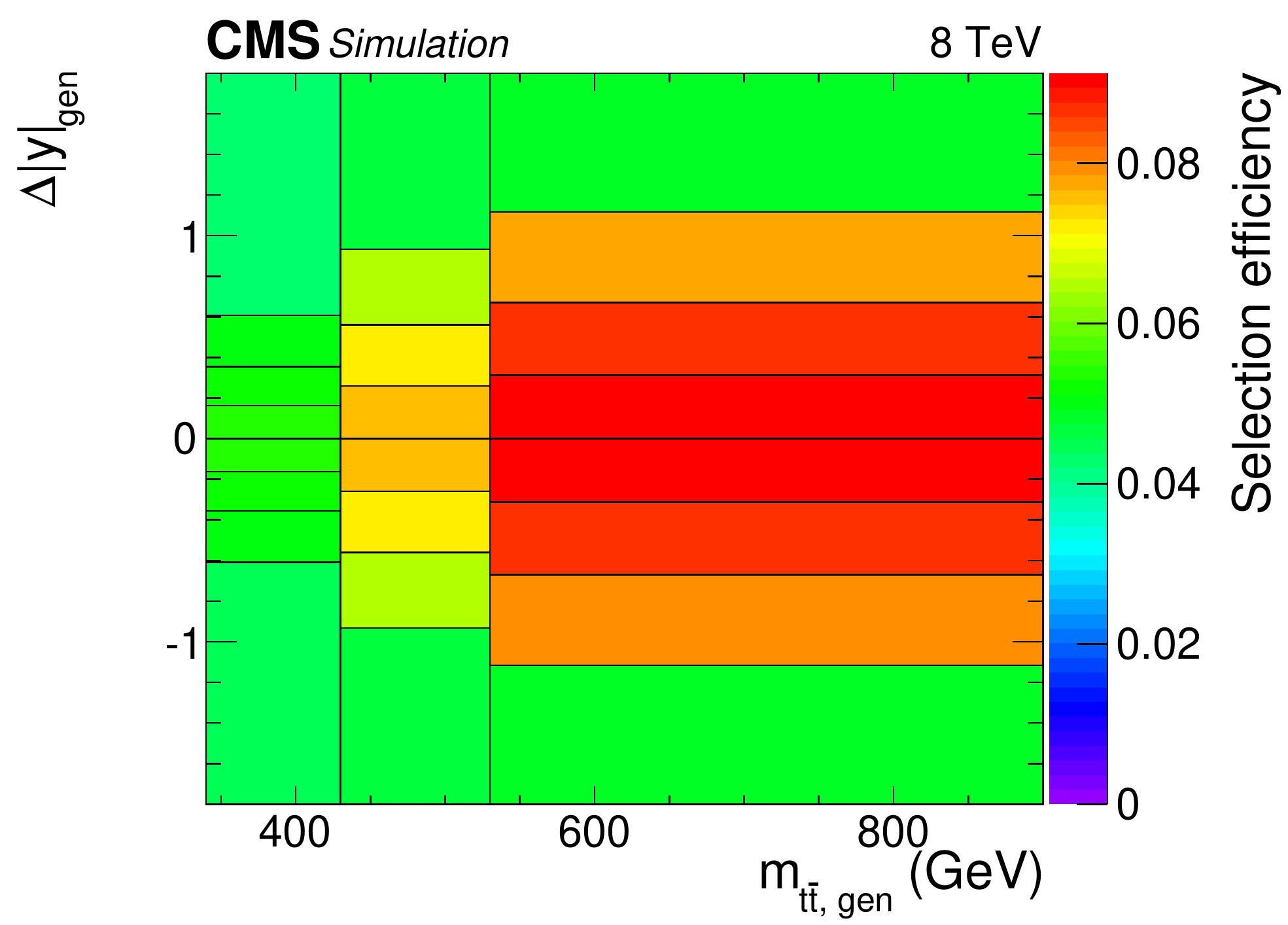}
    \caption{Migration matrix between generated ($\Delta\abs{y}_{\text{gen}}$) and
      reconstructed ($\Delta\abs{y}_{\text{rec}}$) rapidity differences (top left) and selection efficiency
      with respect to the full phase space as a function of $\Delta\abs{y}_{\text{gen}}$ (top right) of the inclusive
      measurement. Migration matrix (bottom left) and selection efficiency
      (bottom right) for the measurement differential in $m_{\ttbar}$.}
 \label{fig:Asy_MigrationMatrices}
\end{figure*}

The resulting background-subtracted distributions are translated from the reconstruction level to parton level within the phase space of the selected events. Afterwards, acceptance corrections are applied, correcting either to the fiducial phase space described in Section~\ref{sec:fidSpaceDef} or to the full phase space. Apart from this last step, the measurements for both phase spaces are identical.
After the corrections have been applied, the resulting distributions are independent of the detector and analysis specifications.

The above corrections are obtained by applying an unfolding procedure to the data~\cite{Blobel:2002pu} through a generalized matrix inversion method.
In this method, the resolution and selection effects are described by a response matrix $R$ that translates the true spectrum $\vec{x}$ into the measured spectrum $\vec{w} = R\vec{x}$.
As reconstruction and selection effects factorize, the response matrix $R$ can be seen as the product of a migration matrix, describing reconstruction effects, and a diagonal matrix containing the selection efficiencies, describing acceptance effects.
Both the migration matrix and the selection efficiencies are determined from simulated \ttbar events.
As the components corresponding to the electron+jets and muon+jets channels are found to be very similar, they are combined to yield a method that can be applied to the summed data of both channels.
In this combination the individual components are scaled according to the scale factors obtained via the background estimation.
The unfolding procedure used in the inclusive measurement, described in detail in Ref.~\cite{Chatrchyan:2011hk}, is also used for the two-dimensional distributions of the differential measurements.

This analysis uses 12 bins for the unfolded $\Delta \abs{y}$ distribution in the inclusive measurement and 8 bins for the same distribution in the differential measurements.
The unfolded $V_{i}$ distributions use 3 bins, with one additional measurement being performed using 6 bins in $m_{\ttbar}$.
The additional measurement provides finely grained results in the region of high $m_{\ttbar}$.
The ranges for the bins in these distributions are given in Table \ref{tab:Binning}.
It should be noted that the outermost bins of $\Delta \abs{y}$ extend to infinity.

\begin{table}[h!tb]
\topcaption{\label{tab:Binning}The bin ranges for the individual bins of the differential measurements. Two different choices of binning are used for the distribution of $m_{\ttbar}$.}
\centering
\begin{tabular}{cr@{--}lr@{--}lr@{--}lr@{--}lr@{--}l}
\hline
Bin& \multicolumn{2}{c}{$\abs{y_{\ttbar}}$} &  \multicolumn{2}{c}{$p_\mathrm{T}^{\ttbar}$ (\GeVns{})\rule{0pt}{12pt}} & \multicolumn{2}{c}{$m_{\ttbar}$ (\GeVns{})}   & \multicolumn{2}{c}{$m_{\ttbar}$ (\GeVns{})} \\  \hline
1& 0    & 0.34        &  0               & 41                &  0   & 430          & 0 & 420 \\
2& 0.34 & 0.75        & \hphantom{.} 41  & 92 \hphantom{.}   &  430 & 530          & 420 & 500  \\
3& 0.75 & $\infty$    &  92              & $\infty$          &  530 & $\infty$     & 500 & 600  \\
4&\multicolumn{2}{c}{}&\multicolumn{2}{c}{}                  & \multicolumn{2}{c}{}& 600 & 750 \\
5&\multicolumn{2}{c}{}&\multicolumn{2}{c}{}                  & \multicolumn{2}{c}{}& 750 & 900 \\
6&\multicolumn{2}{c}{}&\multicolumn{2}{c}{}                  & \multicolumn{2}{c}{}& 900  & $\infty$ \\ \hline
\end{tabular}

\end{table}

In the corresponding reconstructed spectra the numbers of bins along both axes are doubled, as is recommended for the applied unfolding procedure~\cite{Blobel:2002pu}.
The choice of the bin edges for $\Delta \abs{y}$ is different in each bin of $V_{i}$, resulting in different amounts of vertical overlap between horizontally neighbouring bins in the two-dimensional distributions (for illustration see the binning in Fig.~\ref{fig:Asy_MigrationMatrices}, bottom right).

To limit the magnification of statistical uncertainties due to the unfolding procedure, a regularization is applied that suppresses solutions with large fluctuations between neighbouring bins.
The strength of the regularization is determined by minimizing the statistical correlations between bins in the unfolded spectrum. Different strengths are used for the regularization along the sensitive variable within each bin of the kinematic variable. Similarly, the regularization along the kinematic variable is adjusted separately for each bin of the kinematic variable.

Separate migration matrices are used for the inclusive measurement and for each of the differential measurements.
Figure~\ref{fig:Asy_MigrationMatrices} shows the migration matrices for the inclusive measurement and, as an example, for the differential measurement in $m_{\ttbar}$.
For the  inclusive measurement the migration matrix describes the migration of selected events from true values of $\Delta\abs{y}$ to the reconstructed values. For the migration matrices of the differential measurements not only the migration between bins of $\Delta\abs{y}$ has to be taken into account, but also the migration between bins of $V_{i}$.
For a measurement in 3 unfolded bins of $V_{i}$ these migration matrices feature a grid of $6 \times 3$ bins in $V_{i}$, with each of these bins representing a $16 \times 8$ migration matrix describing the migration between different $\Delta\abs{y}$ values.

The values of $\Delta\abs{y}$ and $V_{i}$ also affect the probability for an event to fulfil the event selection criteria.
The selection efficiencies relative to the full phase space for the inclusive measurement and for the differential measurement in $m_{\ttbar}$ are depicted in Fig.~\ref{fig:Asy_MigrationMatrices}. The selection efficiency of the fiducial phase space is defined by the ratio of all selected events to the events present in the fiducial phase space.
It should be noted that the selected events also include events that do not pass the criteria of the fiducial phase space; their influence is implicitly corrected for in the acceptance correction because of the way the selection efficiency is defined.
Thus this correction is multiplicative in nature, which is justified by the inherent similarity of these events to the events that are intended to be measured.

One limiting factor for the precision of the analysis is the presence of sizeable statistical fluctuations in the response matrices as they are obtained from simulated events. To mitigate this effect, one can exploit an approximate symmetry of the response matrix under charge conjugation. For events resulting from a charge-symmetric initial state like gluon-gluon fusion it can be assumed that reconstruction effects also have a predominantly charge-symmetric behaviour. From this reasoning, the symmetry is enforced for this analysis by averaging those bins of the gluon-gluon contribution to the response matrix that correspond to each other under charge conjugation.

The correctness of the unfolding procedure has been verified with pseudo-experiments, each of which provides a randomly generated sample distribution from the templates used in the analysis.

\section{Estimation of systematic uncertainties}
\label{sec:sys}
The measured charge asymmetry $A_{\mathrm{C}}$ is affected by several sources of systematic uncertainty.
Effects altering the direction of the reconstructed top quark momenta can change the value of the reconstructed charge asymmetry.
Systematic uncertainties with an impact on the differential selection efficiency, as well as variations in the rates and modelling of background contributions, can also bias the result.
To evaluate each source of systematic uncertainty, a new background estimation is performed and the measurement is repeated on data using modified simulated samples.
The differences in unfolded asymmetries are then used to construct a systematic asymmetry covariance matrix in a loose analogy to statistical covariance matrices. For an uncertainty described by a single systematic shift a covariance of
\begin{linenomath}
\begin{equation}
 \cov(x,y) = (x-x_\text{nom}) \, (y-y_\text{nom})
\end{equation}
\end{linenomath}
is used, with $x$ and $y$ referring to bins of the asymmetry distribution resulting from the systematic shift and $x_\text{nom}$ and $y_\text{nom}$ being the results of the nominal measurement.
For uncertainties that are determined using exactly two variations (indexed by 1 and 2) the absolute values of the maximal shifts observed in each result bin, $\Delta x_\text{max}$ and $\Delta y_\text{max}$, are determined separately; the covariance is then defined as
\begin{linenomath}
\begin{equation}
 \cov(x,y) = \Delta x_\text{max} \, \Delta y_\text{max} \, \sign\bigl((x_1-x_2) \, (y_1-y_2)\bigr).
\end{equation}
\end{linenomath}
This procedure corresponds to a symmetrization of the largest observed shifts and thus constitutes a more conservative uncertainty estimate than an approach based on a direct analogy with statistical covariance definitions.
The covariance matrices of all systematic uncertainties are added up to yield a resultant matrix where the diagonal elements are the variances.

 In the following, a summary of the studied sources of systematic uncertainty is given.

The corrections to the jet energy scale and jet energy resolution are varied within their $\eta$- and \pt-dependent uncertainties to estimate their effects on the measurement.
 The effect of variations in the frequency of occurrence of pileup events is determined using reweighted simulated samples.
 Differences between data and simulations in the \cPqb~tagging efficiency and the lepton selection efficiency are determined as scale factors that depend on \pt, or $\eta$ and \pt, respectively.
 Effects due to uncertainties on these scale factors are studied by varying them as a function of $\eta$ within their uncertainties and, in the case of the lepton selection efficiency, also as a function of the lepton charge.
 The effect of lepton charge misidentification is very small and is neglected.

 To estimate the influence of a possible mismodelling of the simulated \PW+jets background, the measurement is repeated using a \PW+jets template determined from a sideband region in data, defined by an inversion of the requirement of a b-tagged selected jet.
 The template is reweighted to account for the differences between the signal and sideband regions, which are determined from the simulation.

The uncertainty in the multijet background modelling in the electron+jets channel is determined by replacing the nominal template, which is estimated using two sideband regions defined either by inverted isolation or by inverted identification criteria, with templates derived from only one of the sideband regions each.
 Meanwhile, in the muon+jets channel, only the template from the isolation-inverted sample can be used, so a conservative estimation of the uncertainty in this background contribution is performed by taking the maximum deviation out of three scenarios where the multijet template is replaced with the \ttbar signal template, with the simulated \PW+jets template, or with a template obtained by inverting the sign of the sensitive variable in the multijet template itself.

 In contrast to the other systematic effects, the uncertainty due to the unfolding method is estimated by unfolding simulated samples instead of data.
 The simulated \ttbar events are reweighted to reproduce the observed asymmetries in the differential measurements based on data, and the resulting reconstruction-level spectra are unfolded. The deviations between the unfolded asymmetries and the reweighted true asymmetries are taken to be a measure of the model dependence of the unfolding procedure in the observed point in phase space. The actual uncertainty of each measurement is estimated as the square root of the average squared deviations produced by the unfolding in the three reweighting scenarios corresponding to the three kinematic variables.

To estimate the uncertainty resulting from possible mismodelling of the \ttbar signal, samples of simulated \ttbar events produced with \MADGRAPH are compared to samples produced with \POWHEG, both interfaced to \PYTHIA for the modelling of the parton shower.
 In a similar way the impact of a possible mismodelling of parton shower and hadronization is studied by using \HERWIG~\cite{Corcella:2000bw,Corcella:2002jc}, as opposed to \PYTHIA, for the simulation of the signal, with the hard-scattering matrix element being simulated by either \POWHEG or \MCATNLO~\cite{Frixione:2002ik}.
 As a measure of the uncertainty related to the performed reweighting as a function of the top quark \pt, described in Section~\ref{sec:SimSamples}, the measurement is repeated using samples without reweighting.
  Finally, the impact of variations in the renormalization and factorization scales ($\mu_\mathrm{R}$ and $\mu_\mathrm{F}$) in the simulated \ttbar events is determined using dedicated samples generated at scales varied up and down by factors of 2.

The systematic uncertainty on the measured asymmetry from the choice of PDFs for the colliding protons is estimated using the LHAPDF~\cite{lhapdf} package and the uncertainty in the CT10~\cite{Lai:2010vv}, MSTW2008~\cite{Martin:2009iq}, and NNPDF2.1~\cite{Ball:2011mu} PDF sets.

The contributions of the different sources of systematic uncertainties to the total uncertainty of the inclusive measurements are summarized in Table~\ref{tab:incSyst}.
The table also shows the ranges of systematic uncertainties in the 3-binned differential measurements to illustrate the magnitudes of the individual contributions.
Because the measurements in the two phase spaces differ only by the acceptance corrections, the uncertainties can be seen to behave similarly for the two cases.

 \begin{table*}[thb]
 \topcaption{\label{tab:incSyst}
 Uncertainties for the inclusive measurement of $A_{\mathrm{C}}$ and ranges of uncertainties for the differential measurements using three bins for the kinematic variable. Numbers are given for measurements in the fiducial phase space (fid. PS) and in the full phase space (full PS).}
 \centering
 \begin{tabular}{rccccc} \hline
 Uncertainty source    & \multicolumn{3}{c}{Inclusive $A_{\mathrm{C}}$ uncertainty} & \multicolumn{2}{c}{Differential $A_{\mathrm{C}}$ uncertainty} \\
 &&  fid. PS & full PS  &  fid. PS        & full PS \\ \hline
 Jet energy scale                                && 0.0020   & 0.0018  & 0.0009--0.0066 & 0.0008--0.0063 \\
 Jet energy resolution                           && 0.0003   & 0.0003  & 0.0005--0.0020 & 0.0005--0.0020 \\
 Pileup                                          && 0.0006   & 0.0006  & 0.0002--0.0027 & 0.0003--0.0027 \\
 \cPqb~tagging                                   && 0.0009   & 0.0008  & 0.0002--0.0033 & 0.0002--0.0032 \\
 Lepton selection efficiency                     && 0.0009   & 0.0009  & 0.0005--0.0016 & 0.0005--0.0017 \\
 \PW+jets background                             && 0.0005   & 0.0007  & 0.0003--0.0030 & 0.0005--0.0025 \\
 QCD multijet background                         && 0.0010   & 0.0009  & 0.0008--0.0030 & 0.0011--0.0028 \\
 Unfolding                                       && 0.0012   & 0.0022  & 0.0004--0.0023 & 0.0011--0.0033 \\
 Generator                                       && 0.0002   & 0.0005  & 0.0008--0.0058 & 0.0007--0.0043 \\
 Hadronization                                   && 0.0010   & 0.0011  & 0.0007--0.0046 & 0.0008--0.0040 \\
 Top quark \pt reweighting                       && 0.0000   & 0.0002  & 0.0000--0.0014 & 0.0001--0.0015 \\
 $\mu_\mathrm{R}$ and $\mu_\mathrm{F}$ scales        && 0.0002   & 0.0007  & 0.0008--0.0057 & 0.0009--0.0064 \\
 PDF                                             && 0.0002   & 0.0003  & 0.0004--0.0014 & 0.0004--0.0012 \\ \hline
 Total syst. uncertainty                         && 0.0031   & 0.0037  & 0.0043--0.0120 & 0.0041--0.0115 \\
 Statistical uncertainty                         && 0.0072   & 0.0068  & 0.0078--0.0181 & 0.0078--0.0172 \\\hline
 Total uncertainty                               && 0.0078   & 0.0077  & 0.0094--0.0217 & 0.0094--0.0207 \\\hline
 \end{tabular}
 \end{table*}

\section{Results}

\begin{figure*}[t!hb]
 \centering
   \includegraphics[width=0.49\textwidth]{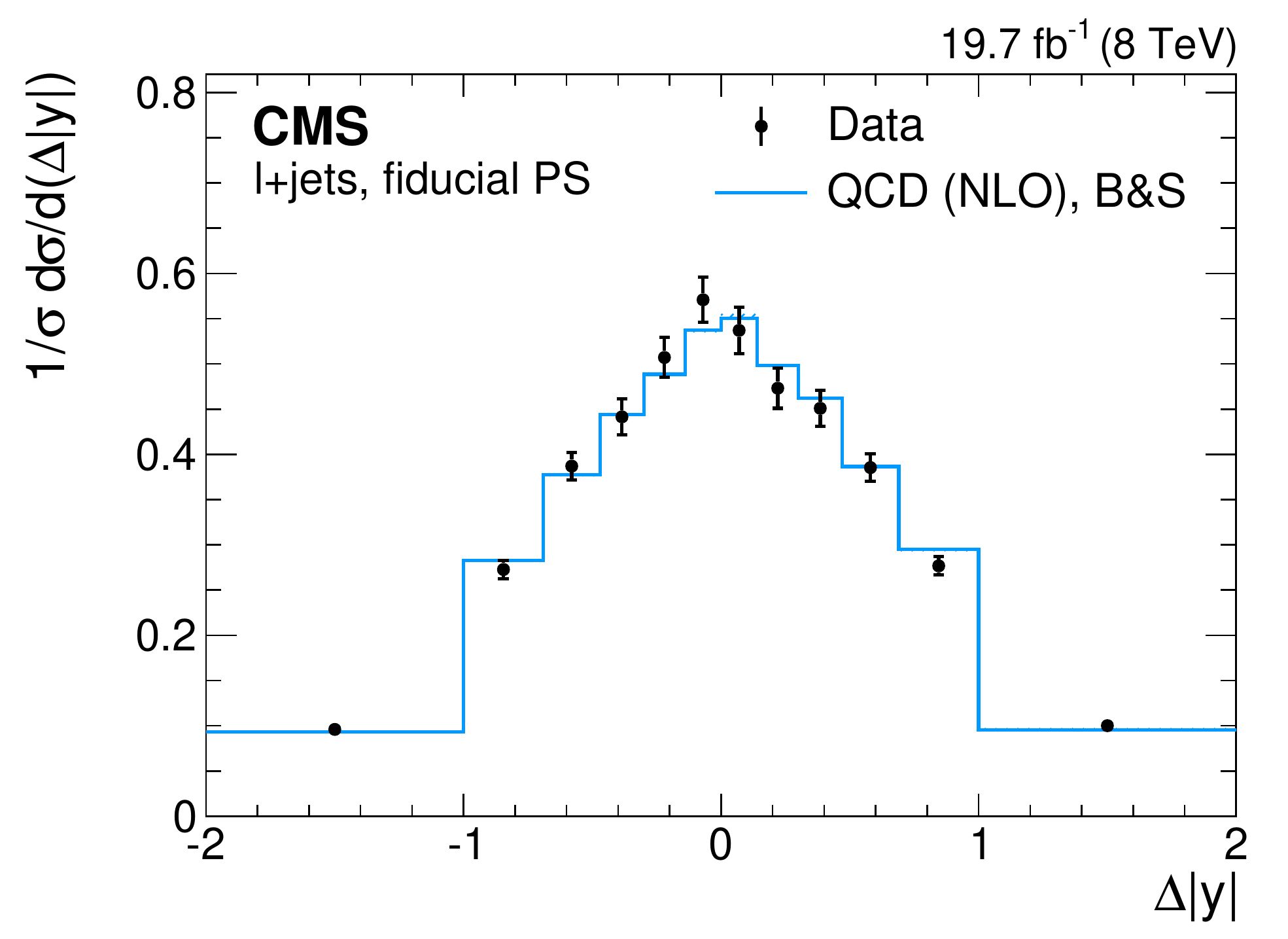}
   \includegraphics[width=0.49\textwidth]{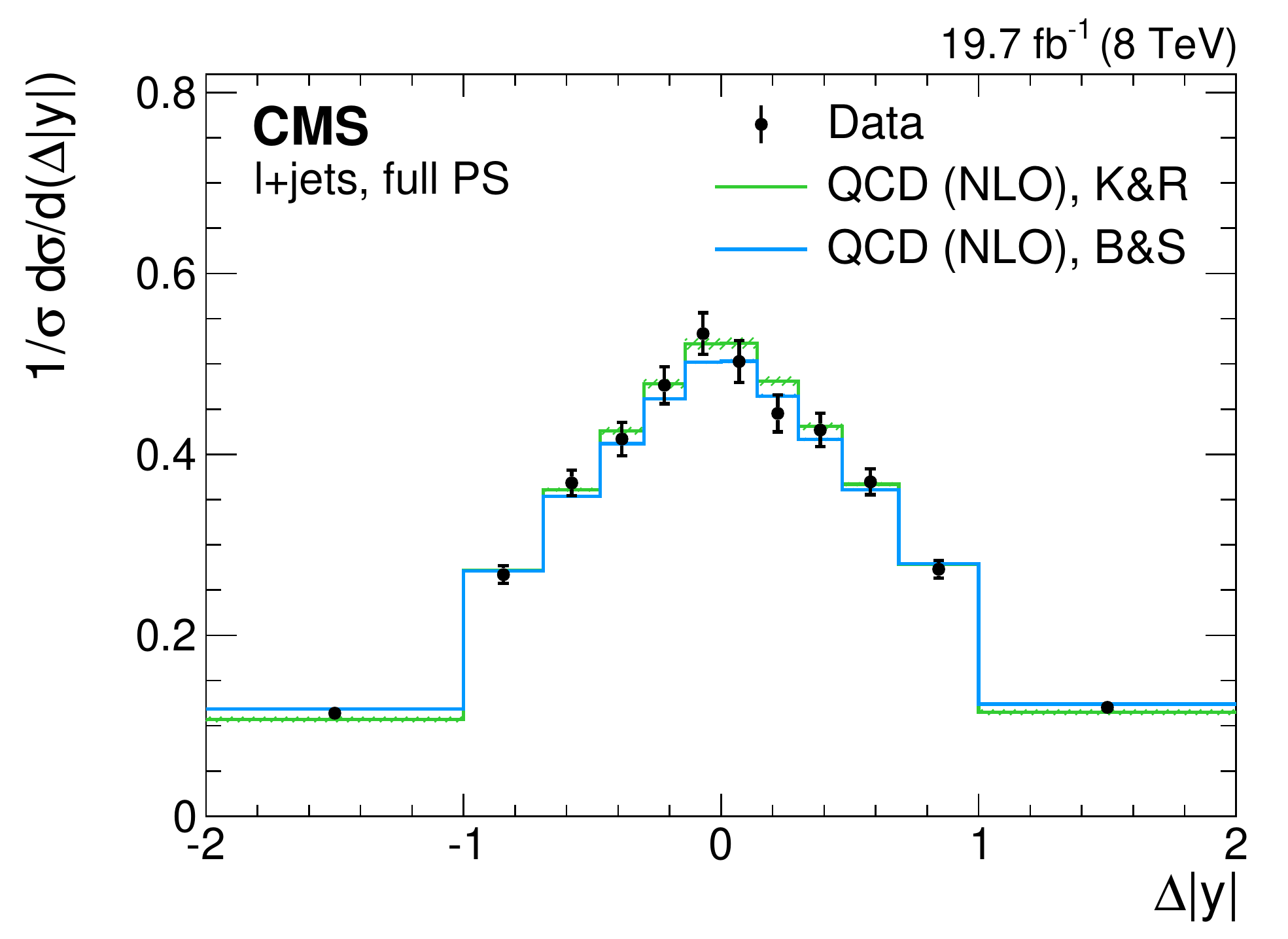}
      \caption{Unfolded inclusive $\Delta\abs{y}$ distribution in the fiducial phase space (left) and in the full phase space (right).
        The uncertainties on the data points represent the statistical uncertainties due to the limited amounts of data and simulated events.
        The measured values are compared to NLO predictions for the SM based on calculations by K\"uhn and Rodrigo (K\&R)~\cite{Kuhn:2011ri} and Bernreuther and Si (B\&S)~\cite{Bernreuther:2012sx,Bernreuther:2010ny}.}
 \label{fig:results1D}
\end{figure*}

Table~\ref{tab:Results_Inc} gives the values of the measured inclusive asymmetry at the different stages of the analysis, while the unfolded $\Delta\abs{y}$ distributions for the fiducial and full phase spaces are shown in Fig.~\ref{fig:results1D}.
The latter two distributions are shown in the form of normalized differential cross sections as a function of $\Delta\abs{y}$.
All unfolded quantities correspond to the parton level. The statistical uncertainty of all quoted results encompasses the subdominant effects of the limited number of simulated events used for the measurement.
It should be noted that the acceptance corrections for the two phase spaces differ as a function of $\Delta\abs{y}$;
as a result, statistical fluctuations in the data can have different effects on the measured asymmetries.

The uncertainty in the theoretical prediction by K\"uhn and Rodrigo~\cite{Kuhn:2011ri} is estimated by varying the top quark mass, the PDFs, and the  $\mu_\textrm{R}$ and $\mu_\textrm{F}$ scales, with the scale uncertainties being the dominant effect.
The uncertainty in the theoretical prediction by Bernreuther and Si~\cite{Bernreuther:2012sx,Bernreuther:2010ny} contains only the effects of variations of the  $\mu_\textrm{R}$ and $\mu_\textrm{F}$ scales.
The \ttbar charge asymmetry for the fiducial phase space is computed with the \ttbar production and semileptonic/non-leptonic \ttbar decay  matrix elements at NLO. The top quark decay matrix elements at NLO contain additional scale dependencies.
This results in a larger scale uncertainty as compared to the charge asymmetry for the full phase space.
Another recent CMS analysis of the inclusive charge asymmetry in the full phase space~\cite{Khachatryan:2015mna}, which uses a slightly more model-dependent approach to achieve lower uncertainties, and a recently published ATLAS measurement of inclusive and differential charge asymmetries~\cite{Aad:2015noh} both yield results that are comparable to the ones presented here.

\begin{table*}[htbp]
\renewcommand{\arraystretch}{1.2}
\topcaption{\label{tab:Results_Inc} The measured inclusive asymmetry at the different stages of the analysis and the corresponding theoretical predictions from the SM.}
 \centering
    \begin{tabular}{lr}
    \hline
                                      &  Asymmetry ($A_{\mathrm{C}}$) \\
    \hline
    Reconstructed                     & $0.0036 \pm 0.0017\stat$\\
    Background-subtracted                     & $0.0008 \pm 0.0023\stat$\\
    Corrected for migration effects   & $-0.0042 \pm 0.0072\stat$\\
    \hline
    Fiducial phase space              & $-0.0035 \pm 0.0072\stat\pm 0.0031\syst$\\
    Theoretical prediction  [Bernreuther, Si]~\cite{Bernreuther:2012sx,Bernreuther:2010ny} &$0.0101 \pm 0.0010 $  \\
    \hline
    Full phase space                  &$ 0.0010 \pm 0.0068\stat\pm 0.0037\syst$\\
    Theoretical prediction  [K\"uhn, Rodrigo]~\cite{Kuhn:2011ri}  &$0.0102 \pm 0.0005 $  \\
    Theoretical prediction  [Bernreuther, Si]~\cite{Bernreuther:2012sx,Bernreuther:2010ny} &$0.0111 \pm 0.0004 $  \\
    \hline

 \end{tabular}
\end{table*}

\begin{figure*}[th!]
\centering
\includegraphics[width=0.49\textwidth]{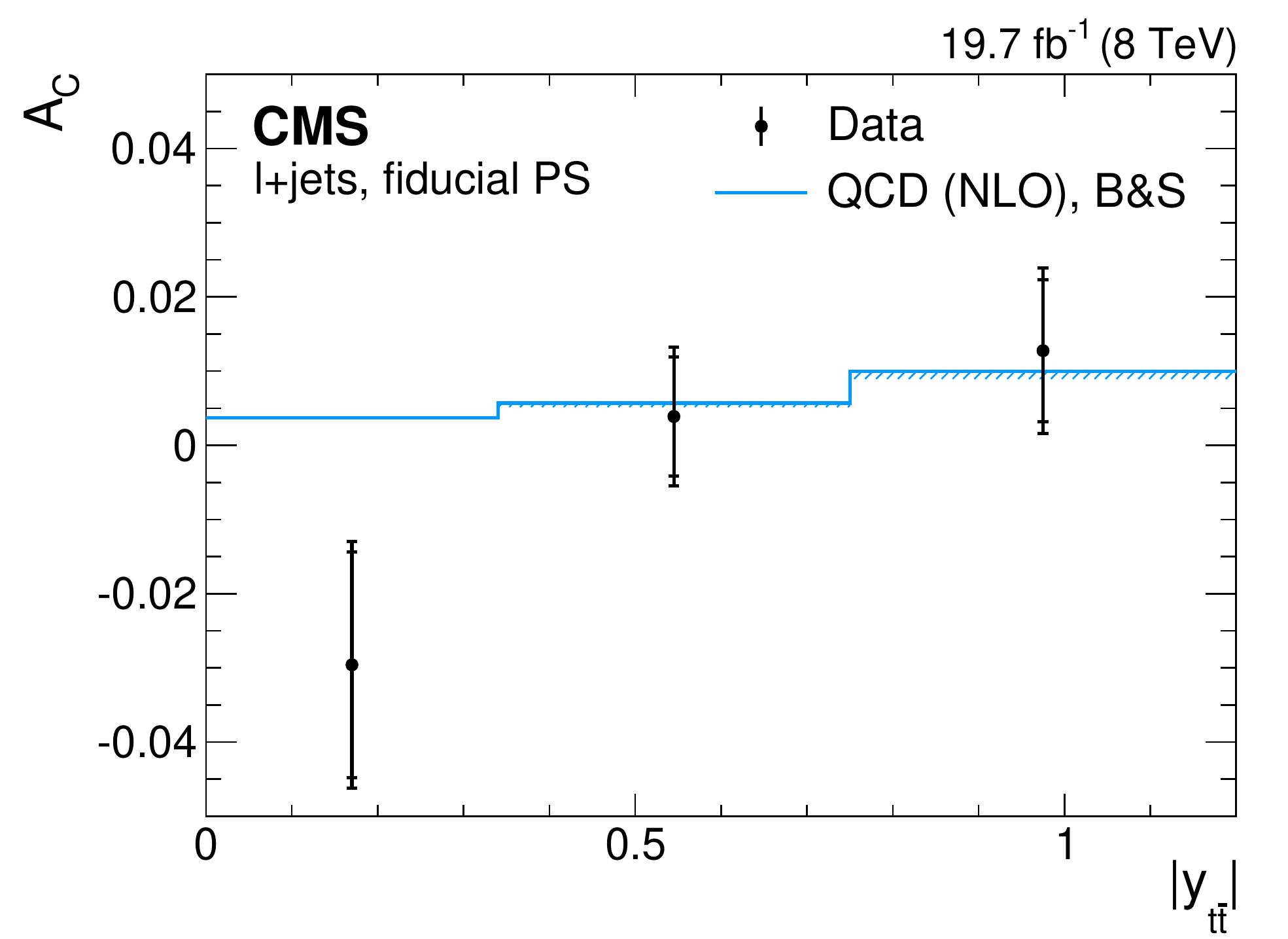}
\includegraphics[width=0.49\textwidth]{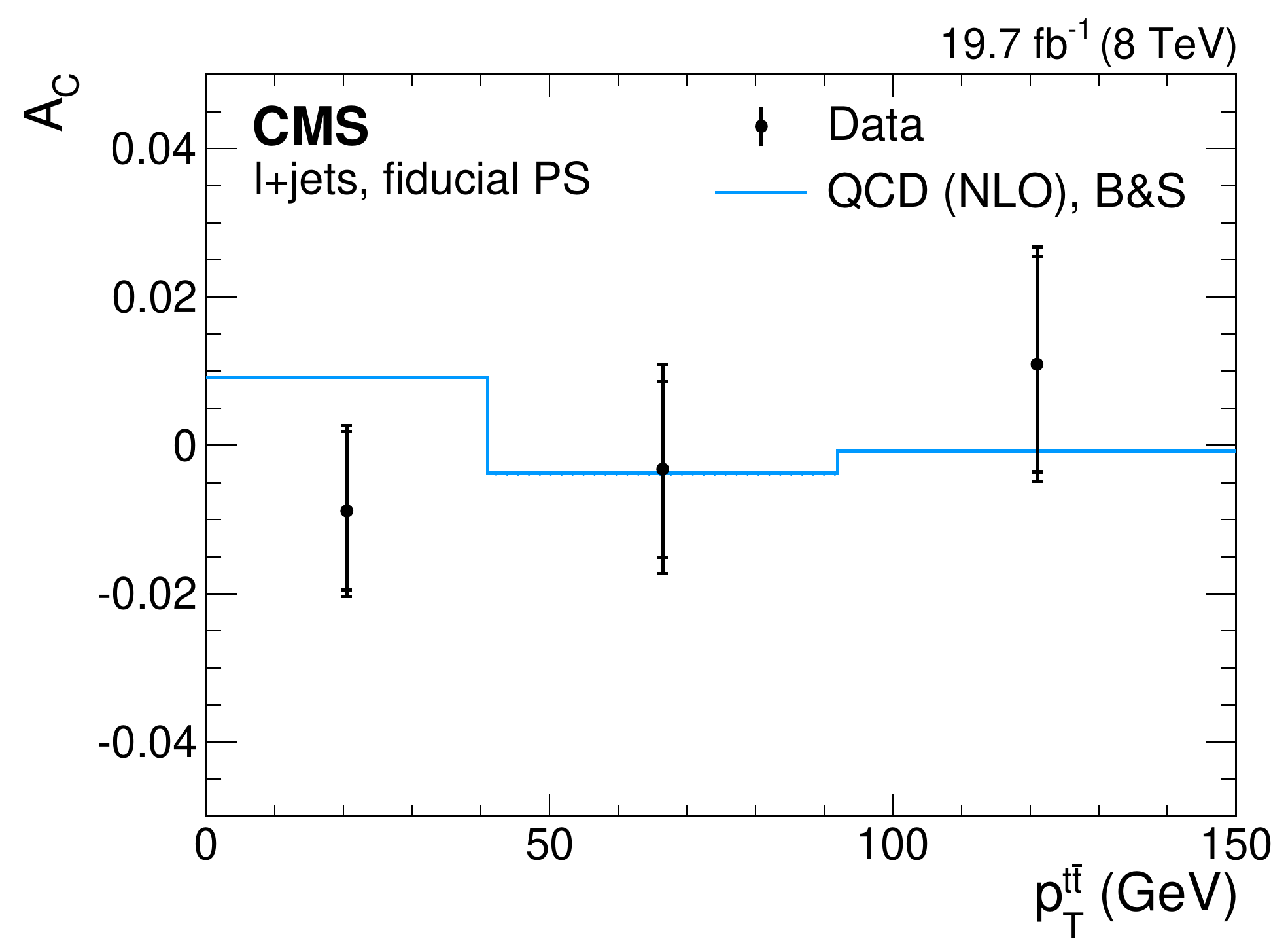}
\includegraphics[width=0.49\textwidth]{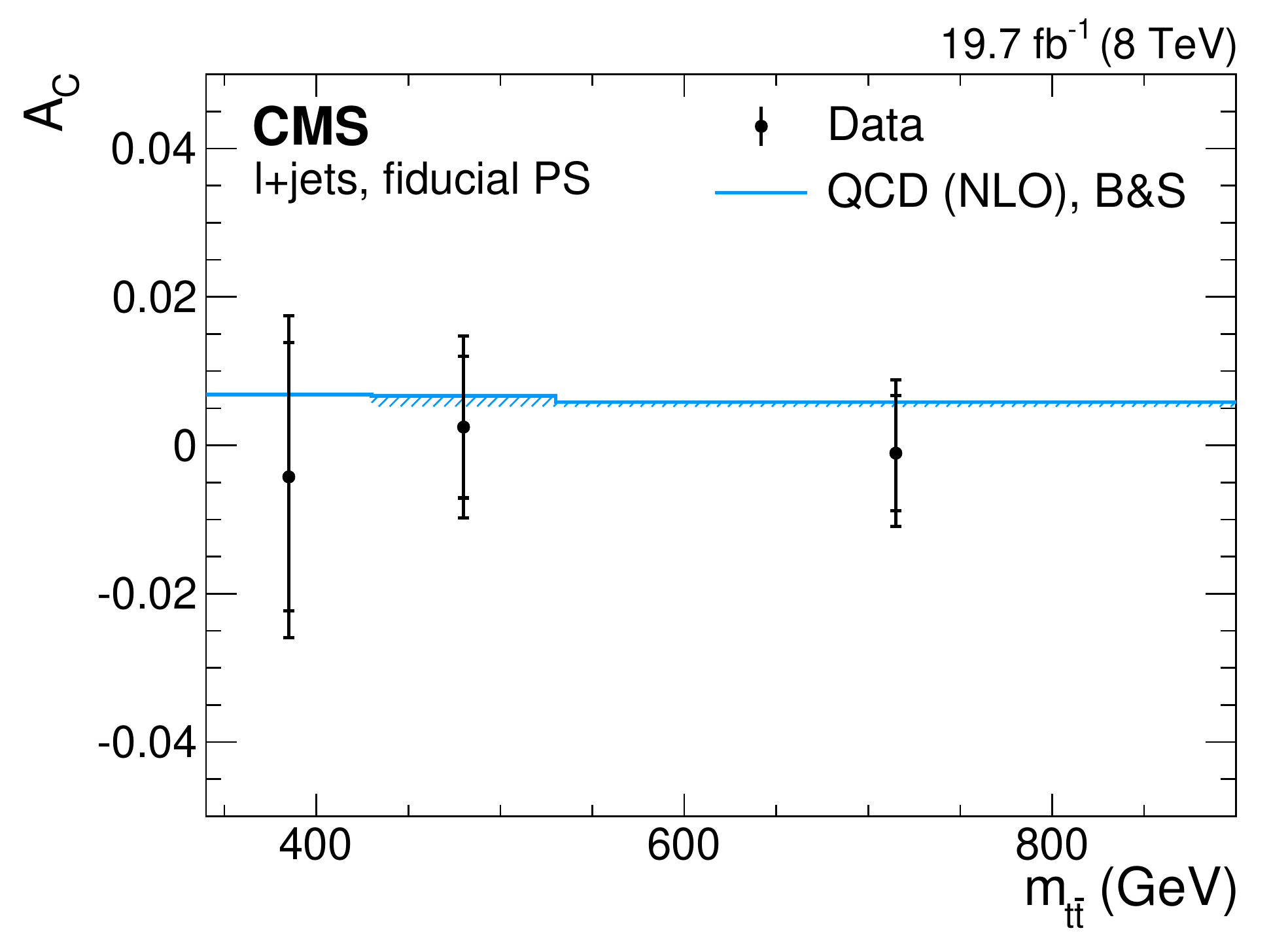}
\includegraphics[width=0.49\textwidth]{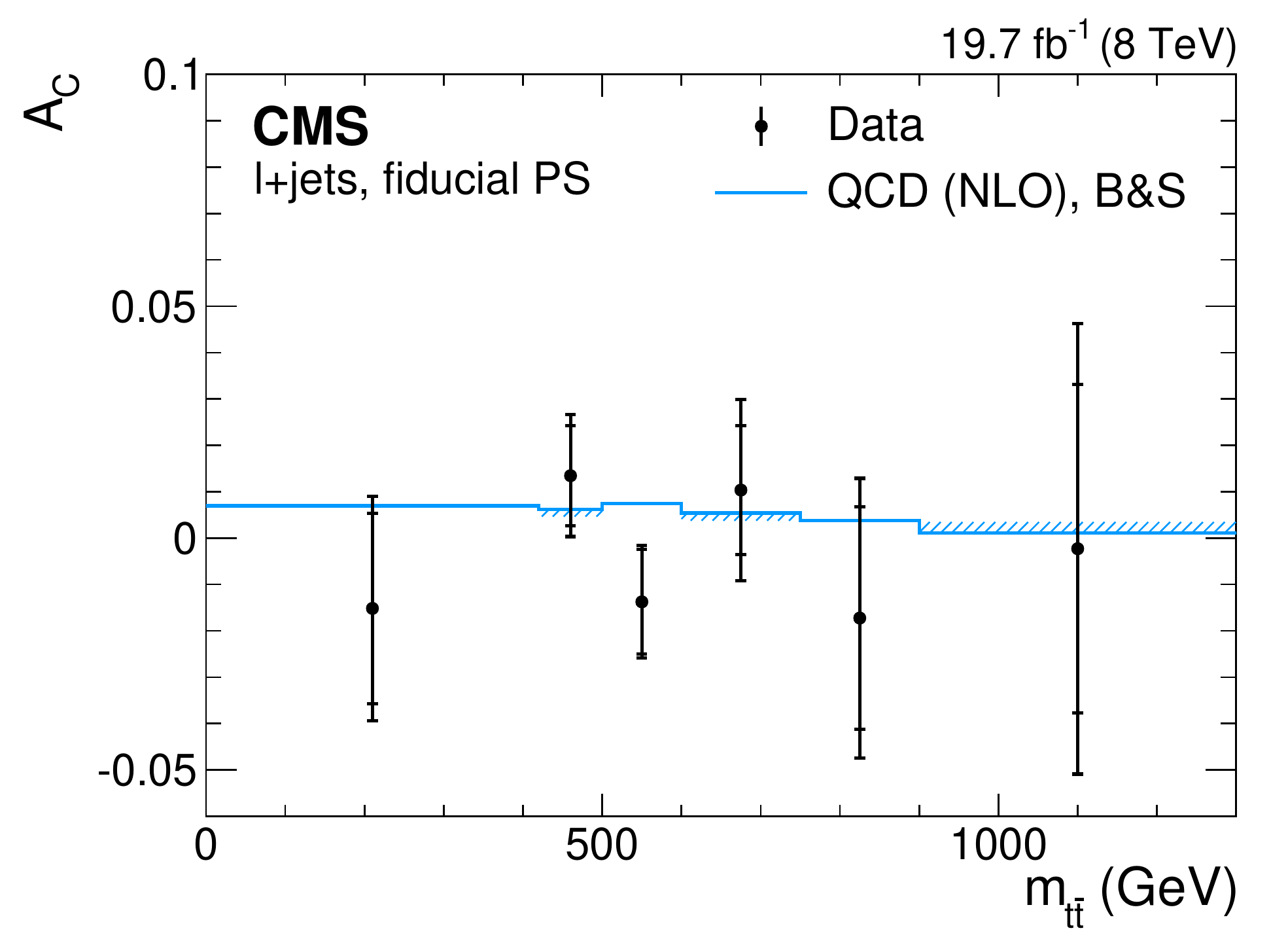}
\caption{Corrected asymmetry as a function of $\abs{y_{\ttbar}}$ (upper left), $p_\mathrm{T}^{\ttbar}$ (upper right), and $m_{\ttbar}$ (lower left and lower right). The latter is shown in two different binnings. All results correspond to the \emph{fiducial phase space}.
The measured values are compared to an NLO prediction for the SM based on calculations by Bernreuther and Si (B\&S)~\cite{Bernreuther:2012sx,Bernreuther:2010ny}. The inner bars indicate the statistical uncertainties, while the outer bars represent the statistical and systematic uncertainties added in quadrature.}
\label{fig:results_FidParticle}
\end{figure*}

\begin{figure*}[th!]
 \centering
   \includegraphics[width=0.49\textwidth]{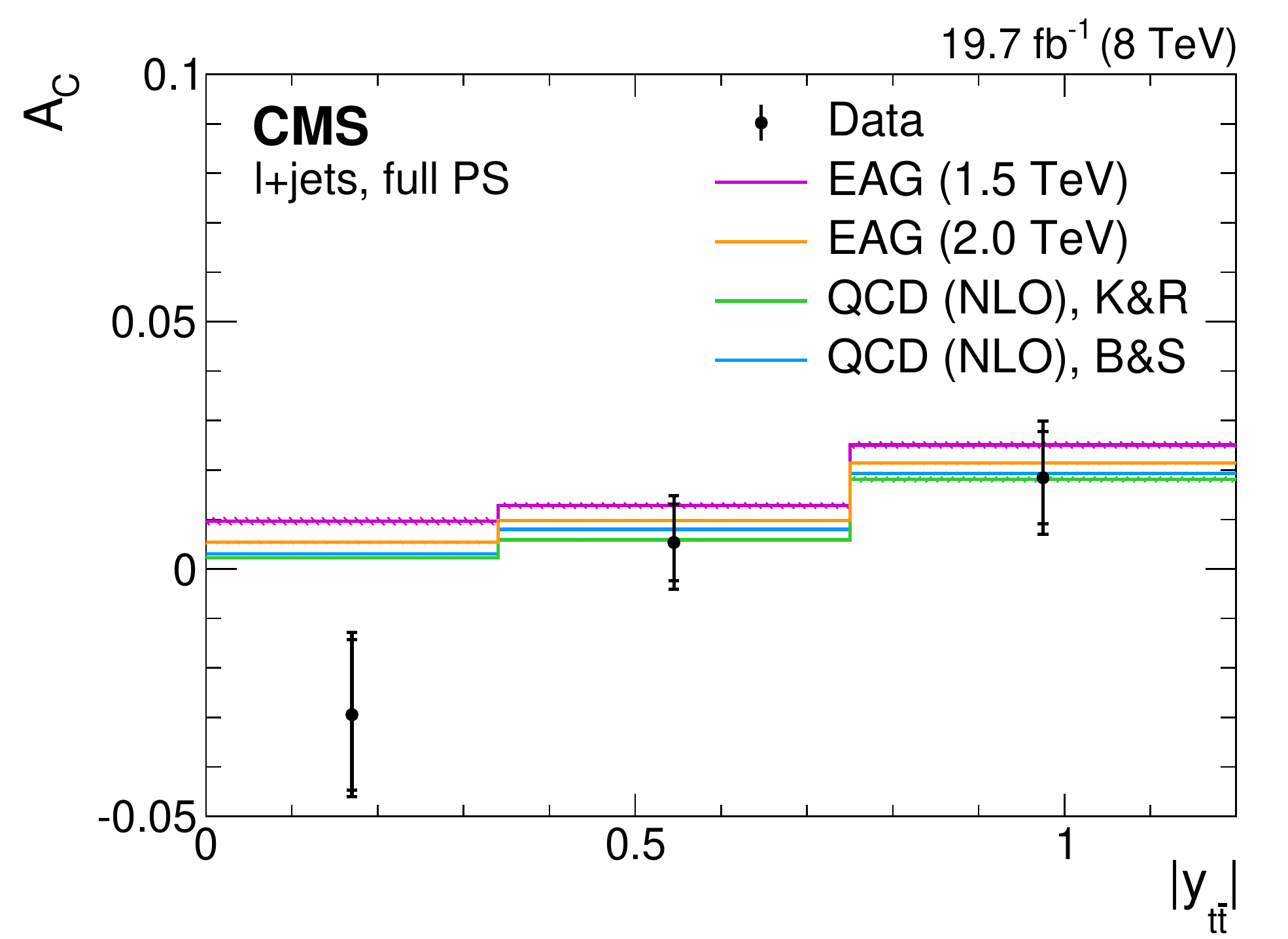}
   \includegraphics[width=0.49\textwidth]{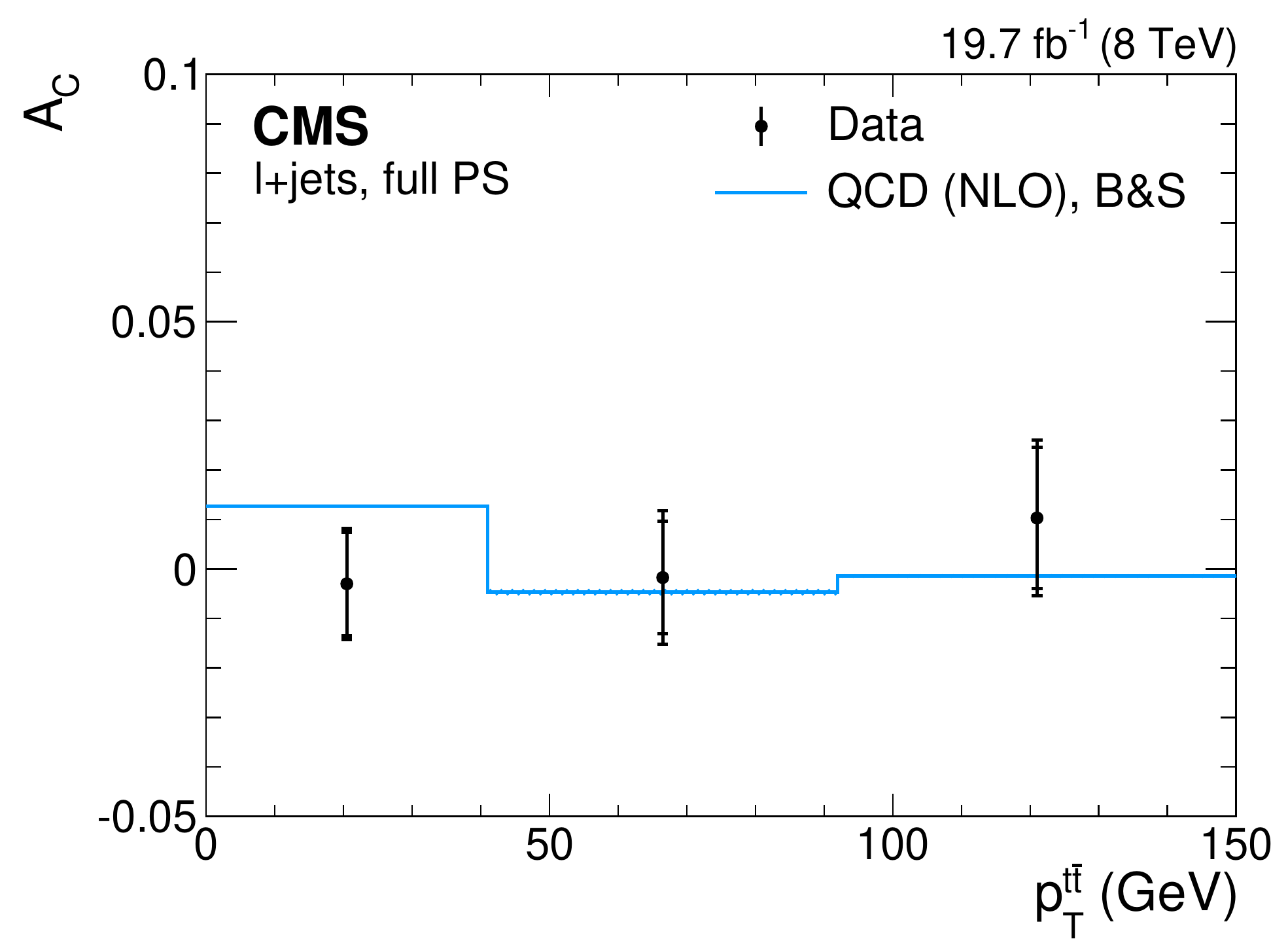}
   \includegraphics[width=0.49\textwidth]{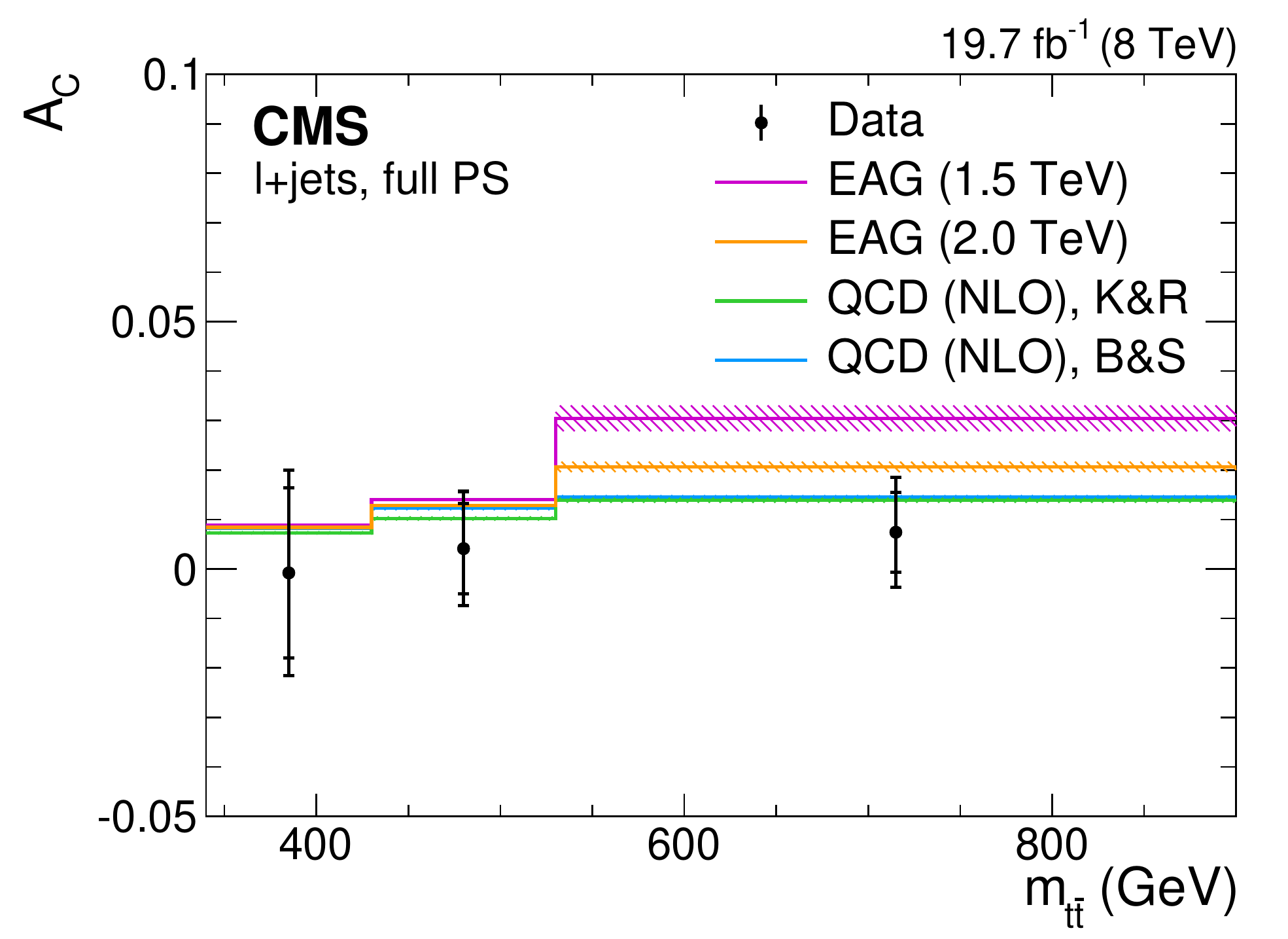}
   \includegraphics[width=0.49\textwidth]{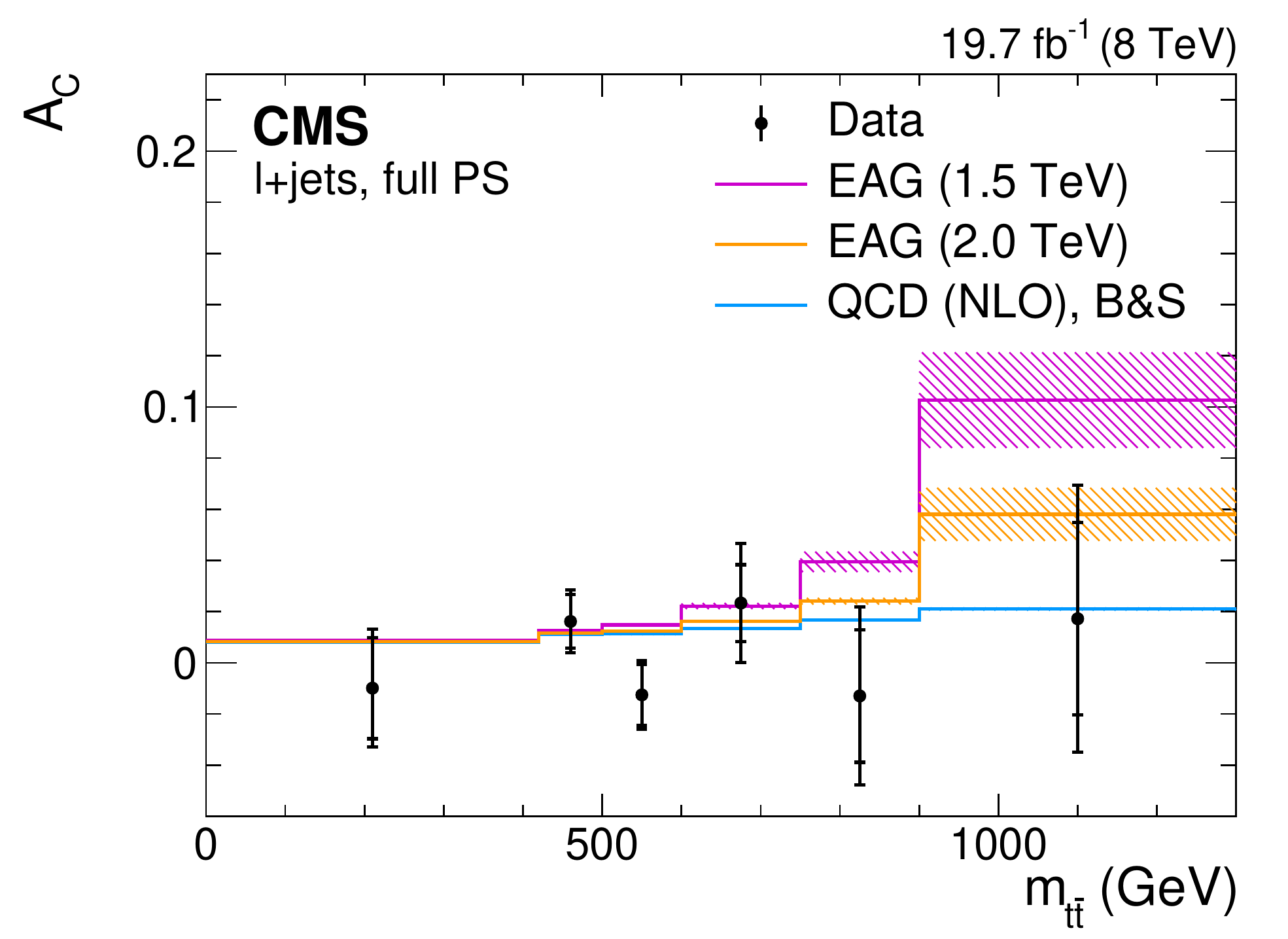}
   \caption{Corrected asymmetry as a function of $\abs{y_{\ttbar}}$ (upper left), $p_\mathrm{T}^{\ttbar}$ (upper right), and $m_{\ttbar}$ (lower left and lower right). The latter is shown in two different binnings. All results correspond to the \emph{full phase space}.
   The measured values are compared to NLO predictions for the SM based on calculations by K\"uhn and Rodrigo (K\&R)~\cite{Kuhn:2011ri} and Bernreuther and Si (B\&S)~\cite{Bernreuther:2012sx,Bernreuther:2010ny}, as well as to the predictions of a model featuring an effective axial-vector coupling of the gluon (EAG)~\cite{Gabrielli:2011zw,Gabrielli:2011jf}.
     The inner bars indicate the statistical uncertainties, while the outer bars represent the statistical and systematic uncertainties added in quadrature.}
 \label{fig:results2D}
\end{figure*}

The results of the differential measurements in the fiducial phase space are shown in Fig.~\ref{fig:results_FidParticle}, and the extrapolation to the full phase space in Fig.~\ref{fig:results2D}.
The measured values are compared to predictions from SM
calculations~\cite{Kuhn:2011ri, Bernreuther:2012sx,Bernreuther:2010ny} as well as to predictions from an effective field theory~\cite{Gabrielli:2011zw,Gabrielli:2011jf}.
The latter is capable of reproducing the CDF results~\cite{Aaltonen:2012it} by introducing an anomalous effective axial-vector coupling to the gluon at the one-loop level.
The gluon-quark vertex is treated in the approximation of an effective field theory with a scale for new physics contributions of order 1.5--2.0\TeV.
Predictions for the asymmetry as a function of $\pt^{\ttbar}$ are not available for this theory and for one of the SM calculations.
Because of the importance of the region of high $m_{\ttbar}$ for the detection of new physics, we provide an additional, more finely-grained differential measurement of the asymmetry as a function of this observable.

Both of the inclusive results yield values that are slightly smaller than the SM predictions, with the larger deviation being in the fiducial phase space and corresponding to 1.7 standard deviations.
The differential measurements show a good agreement with the SM predictions.
For the benchmark model involving an effective axial-vector coupling of the gluon, the measurement at high $m_{\ttbar}$ excludes new physics scales below $1.5\TeV$ at the 95\% confidence level.

\section{Summary}
Inclusive and differential measurements of the charge asymmetry in \ttbar production at the LHC are presented.
The data sample, collected in proton-proton collisions at $\sqrt{s}=8\TeV$ with the CMS detector, corresponds to an integrated luminosity of 19.7\fbinv.
Events with top quark pairs decaying into the electron+jets and muon+jets channels are selected and a full \ttbar event reconstruction is performed to determine the four-momenta of the top quarks and antiquarks.
The observed distributions are then corrected for acceptance and reconstruction effects.
For the first time at the LHC, acceptance corrections to the \ttbar charge asymmetry are performed not only to the full phase space but also to a fiducial phase space.
Within two standard deviations, all measured values are consistent with the predictions of the standard model and no hint of new physics contributions is observed.
The charge asymmetry in the high-mass region is about two standard deviations below the predictions from an effective field theory with the scale for new physics at 1.5\TeV.

\begin{acknowledgments}
We thank Werner Bernreuther, Zong-Guo Si, Johann K\"uhn, German Rodrigo, Emidio Gabrielli, Antonio Racioppi, and Martti Raidal for kindly providing theoretical predictions for the differential distributions that are measured in this letter.

We congratulate our colleagues in the CERN accelerator departments for the excellent performance of the LHC and thank the technical and administrative staffs at CERN and at other CMS institutes for their contributions to the success of the CMS effort. In addition, we gratefully acknowledge the computing centres and personnel of the Worldwide LHC Computing Grid for delivering so effectively the computing infrastructure essential to our analyses. Finally, we acknowledge the enduring support for the construction and operation of the LHC and the CMS detector provided by the following funding agencies: BMWFW and FWF (Austria); FNRS and FWO (Belgium); CNPq, CAPES, FAPERJ, and FAPESP (Brazil); MES (Bulgaria); CERN; CAS, MoST, and NSFC (China); COLCIENCIAS (Colombia); MSES and CSF (Croatia); RPF (Cyprus); MoER, ERC IUT and ERDF (Estonia); Academy of Finland, MEC, and HIP (Finland); CEA and CNRS/IN2P3 (France); BMBF, DFG, and HGF (Germany); GSRT (Greece); OTKA and NIH (Hungary); DAE and DST (India); IPM (Iran); SFI (Ireland); INFN (Italy); MSIP and NRF (Republic of Korea); LAS (Lithuania); MOE and UM (Malaysia); CINVESTAV, CONACYT, SEP, and UASLP-FAI (Mexico); MBIE (New Zealand); PAEC (Pakistan); MSHE and NSC (Poland); FCT (Portugal); JINR (Dubna); MON, RosAtom, RAS and RFBR (Russia); MESTD (Serbia); SEIDI and CPAN (Spain); Swiss Funding Agencies (Switzerland); MST (Taipei); ThEPCenter, IPST, STAR and NSTDA (Thailand); TUBITAK and TAEK (Turkey); NASU and SFFR (Ukraine); STFC (United Kingdom); DOE and NSF (USA).

Individuals have received support from the Marie-Curie programme and the European Research Council and EPLANET (European Union); the Leventis Foundation; the A. P. Sloan Foundation; the Alexander von Humboldt Foundation; the Belgian Federal Science Policy Office; the Fonds pour la Formation \`a la Recherche dans l'Industrie et dans l'Agriculture (FRIA-Belgium); the Agentschap voor Innovatie door Wetenschap en Technologie (IWT-Belgium); the Ministry of Education, Youth and Sports (MEYS) of the Czech Republic; the Council of Science and Industrial Research, India; the HOMING PLUS programme of the Foundation for Polish Science, cofinanced from European Union, Regional Development Fund; the Compagnia di San Paolo (Torino); the Consorzio per la Fisica (Trieste); MIUR project 20108T4XTM (Italy); the Thalis and Aristeia programmes cofinanced by EU-ESF and the Greek NSRF; the National Priorities Research Program by Qatar National Research Fund; the Rachadapisek Sompot Fund for Postdoctoral Fellowship, Chulalongkorn University (Thailand); and the Welch Foundation.
\end{acknowledgments}

\bibliography{auto_generated}

\cleardoublepage \appendix\section{The CMS Collaboration \label{app:collab}}\begin{sloppypar}\hyphenpenalty=5000\widowpenalty=500\clubpenalty=5000\textbf{Yerevan Physics Institute,  Yerevan,  Armenia}\\*[0pt]
V.~Khachatryan, A.M.~Sirunyan, A.~Tumasyan
\vskip\cmsinstskip
\textbf{Institut f\"{u}r Hochenergiephysik der OeAW,  Wien,  Austria}\\*[0pt]
W.~Adam, E.~Asilar, T.~Bergauer, J.~Brandstetter, E.~Brondolin, M.~Dragicevic, J.~Er\"{o}, M.~Flechl, M.~Friedl, R.~Fr\"{u}hwirth\cmsAuthorMark{1}, V.M.~Ghete, C.~Hartl, N.~H\"{o}rmann, J.~Hrubec, M.~Jeitler\cmsAuthorMark{1}, V.~Kn\"{u}nz, A.~K\"{o}nig, M.~Krammer\cmsAuthorMark{1}, I.~Kr\"{a}tschmer, D.~Liko, T.~Matsushita, I.~Mikulec, D.~Rabady\cmsAuthorMark{2}, B.~Rahbaran, H.~Rohringer, J.~Schieck\cmsAuthorMark{1}, R.~Sch\"{o}fbeck, J.~Strauss, W.~Treberer-Treberspurg, W.~Waltenberger, C.-E.~Wulz\cmsAuthorMark{1}
% ...
\vskip\cmsinstskip
% ...
1:~~Also at Vienna University of Technology, Vienna, Austria\\
2:~~Also at CERN, European Organization for Nuclear Research, Geneva, Switzerland\\
% ...

\end{sloppypar}
\end{document}